\documentclass[a4paper,fleqn]{cas-dc}
\usepackage[utf8]{inputenc}
\usepackage[numbers]{natbib}
\usepackage[T1]{fontenc}
\usepackage{lineno}
\usepackage{hyphenat}
\modulolinenumbers[5]

\usepackage[dvipsnames]{xcolor}
\usepackage{xspace}

\usepackage{csvsimple}
\usepackage{longtable}
\usepackage{lscape}
\usepackage{array}
\usepackage{subfig}
\usepackage{array, etoolbox}

\usepackage[linesnumbered]{algorithm2e}

\usepackage{color}
\definecolor{pblue}{rgb}{0.13,0.13,1}
\definecolor{pgreen}{rgb}{0,0.5,0}
\definecolor{pred}{rgb}{0.9,0,0}
\definecolor{pgrey}{rgb}{0.46,0.45,0.48}

\usepackage[scaled]{beramono}
\usepackage{tikz}
\usepackage{xspace}
\usepackage{listings}
\makeatletter
\newenvironment{btHighlight}[1][]
{\begingroup\tikzset{bt@Highlight@par/.style={#1}}\begin{lrbox}{\@tempboxa}}
{\end{lrbox}\bt@HL@box[bt@Highlight@par]{\@tempboxa}\endgroup}
\newcommand\btHL[1][]{%
  \begin{btHighlight}[#1]\bgroup\aftergroup\bt@HL@endenv%
}
\def\bt@HL@endenv{%
  \end{btHighlight}%
  \egroup
}
\newcommand{\bt@HL@box}[2][]{%
  \tikz[#1]{%
    \pgfpathrectangle{\pgfpoint{1pt}{0pt}}{\pgfpoint{\wd #2}{\ht #2}}%
    \pgfusepath{use as bounding box}%
    \node[anchor=base west, fill=orange!30,outer sep=0pt,inner xsep=0pt, inner ysep=0pt, minimum height=\ht\strutbox+1pt,#1]{\raisebox{1pt}{\strut}\strut\usebox{#2}};
  }%
}
\lstdefinestyle{Java}{
    language={Java}, basicstyle=\ttfamily\footnotesize, 
    moredelim=**[is][{\btHL[fill=red!17,thin]}]{`}{`},
    moredelim=**[is][{\btHL[fill=green!17,thin]}]{£}{£},
    moredelim=**[is][{\btHL[fill=yellow!17,thin]}]{~}{~},
    moredelim=**[is][{\color{green!17}\btHL[fill=green!17,thin]}]{¤}{¤},
    moredelim=**[is][{\color{red!17}\btHL[fill=red!17,thin]}]{µ}{µ},
}

\usepackage{dsfont}
\usepackage{lineno}
\usepackage[T1]{fontenc}
\usepackage[flushleft]{threeparttable}
\usepackage{multicol}
\usepackage{xspace}
\usepackage{amssymb}
\usepackage{amsthm}
\usepackage{lscape}
\usepackage{multirow}
\usepackage{array}
\usepackage{paralist}
\usepackage{makecell}
\usepackage{graphicx}
\usepackage{tabu}
\usepackage[framemethod=tikz]{mdframed}
\newtheorem{defn}{Definition}

\mdfdefinestyle{mpdframe}{
    frametitlebackgroundcolor   =black!15,
    frametitlerule              =true,
    roundcorner                 =5pt,
    middlelinewidth             =0.8pt,
    innermargin                 =0.2cm,
    outermargin                 =0.2cm,
    innerleftmargin             =0.2cm,
    innerrightmargin            =0.2cm,
    innertopmargin              =0.2cm,
    innerbottommargin           =0.2cm
}

\usepackage[colorinlistoftodos,prependcaption]{todonotes}
\newboolean{showcomments}
\setboolean{showcomments}{true}
\ifthenelse{\boolean{showcomments}}
 { \newcommand{\mynote}[2]{
      \fbox{\bfseries\sffamily\scriptsize#1}
        {\small$\blacktriangleright$\textsf{\textcolor{red}{{\em #2}\bf }}$\blacktriangleleft$}}}
        { \newcommand{\mynote}[2]{}}
\usepackage{tabularx}

\usepackage{booktabs}

\definecolor{mymauve}{rgb}{0.58,0,0.82}
\definecolor{mygrey}{rgb}{0.43, 0.5, 0.5}

\usepackage{pbox}

\input{utils/util.tex}

\newcounter{rowcount}
\usepackage{amsthm}
\theoremstyle{definition}

\usepackage[framemethod=tikz]{mdframed}
\mdfdefinestyle{mpdframe}{
    frametitlebackgroundcolor   =black!15,
    frametitlerule              =true,
    roundcorner                 =5pt,
    middlelinewidth             =1pt,
    innermargin                 =0.2cm,
    outermargin                 =0.2cm,
    innerleftmargin             =0.2cm,
    innerrightmargin            =0.2cm,
    innertopmargin              =0.2cm,
    innerbottommargin           =0.2cm
}

\usepackage{hyperref}
\usepackage{url}
\AtBeginDocument{%
}

\begin{document}

\hyphenation{de-compi-ler}
\hyphenation{ba-sed}
\hyphenation{meta-de-com-pi-la-tion}

\let\WriteBookmarks\relax
\def\floatpagepagefraction{1}
\def\textpagefraction{.001}
\newcommand\mytitle{Java Decompiler Diversity and its Application to Meta-decompilation}
\shorttitle{\mytitle}
\shortauthors{Harrand et~al.}

\title [mode = title]{\mytitle}

\author{Nicolas Harrand}[orcid=0000-0002-2491-2771]\ead{harrand@kth.se}\cormark[1]

\author{C\'esar Soto-Valero}[orcid=0000-0003-0541-6411]\ead{cesarsv@kth.se}

\author{Martin Monperrus}[orcid=0000-0003-3505-3383]\ead{martin.monperrus@csc.kth.se}

\author{Benoit Baudry}[orcid=0000-0002-4015-4640]\ead{baudry@kth.se}

\cortext[cor1]{Corresponding author}

\address[]{KTH Royal Institute of Technology, SE-100 44 Stockholm, Sweden}

\begin{abstract}
During compilation from Java source code to bytecode, some information is irreversibly lost. In other words, compilation and decompilation of Java code is not symmetric. Consequently, decompilation, which aims at producing source code from bytecode, relies on strategies to reconstruct the information that has been lost. Different Java decompilers  use distinct strategies to achieve proper decompilation. In this work, we hypothesize that the diverse ways in which bytecode can be decompiled has a direct impact on the quality of the source code produced by decompilers. 

In this paper, we assess the strategies of eight Java decompilers with respect to three quality indicators: syntactic correctness, syntactic distortion and \semi.
Our results show that no single modern decompiler is able to correctly handle the variety of bytecode structures coming from real-world programs. The highest ranking decompiler in this study produces syntactically correct, and semantically equivalent code output for 84\%, respectively 78\%, of the classes in our dataset.
Our results demonstrate that each decompiler correctly handles a different set of bytecode classes.

We  propose a new decompiler called Arlecchino that leverages the diversity of existing decompilers. To do so, we merge partial decompilation into a new one based on compilation errors. Arlecchino handles $37.6\%$ of bytecode classes that were previously handled by no decompiler. We publish the sources of this new bytecode decompiler.
\end{abstract}


\newcommand\hhighlights{\begin{itemize}
    \item an empirical comparison of eight Java decompilers based on $2041$ real-world Java classes, tested by $25019$ test cases, identifying the key strengths and limitations of Java bytecode decompilation; 
    \item meta-decompilation, a novel approach to decompilation that leverages decompilers diversity to improve decompilation effectiveness;
    \item a tool and a dataset for future research on Java decompilers publicly available at \url{https://github.com/castor-software/decompilercmp}
\end{itemize}}


\maketitle

\section{Introduction}\label{sec:introduction}

In the Java programming language, source code is compiled into an intermediate stack-based representation known as bytecode, which is  interpreted by the Java Virtual Machine (JVM). In the process of translating source code to bytecode, the compiler performs various analyses. Even if most optimizations are typically performed at runtime by the just-in-time (JIT) compiler, several pieces of information residing in the original source code are already not present in the bytecode anymore due to compiler optimization~\cite{Miecznikowski2002,Lindholm2014}. For example the structure of loops is altered and local variable names may be modified~\cite{Jaffe2018}.

Decompilation is the inverse process, it consists in transforming the  bytecode instructions into source code~\cite{Nolan2004}. 
Decompilation can be done with several goals in mind.
First, it can be used to help developers understand the code of the libraries they use. This is why Java IDEs such as IntelliJ and Eclipse include built-in decompilers to help developers analyze the third-party classes for which the source code is not available. In this case, the readability of the decompiled code is paramount.
Second, decompilation may be a preliminary step before another compilation pass, for example with a different compiler. In this case, the main goal is that the decompiled code is syntactically correct and can be recompiled.
Some other applications of decompilation with slightly different criteria include clone detection~\cite{Ragkhitwetsagul2017}, malware analysis ~\cite{Yakdan2016, Durfina2013} and software archaeology~\cite{Robles2005}.\looseness=-1

Overall, the ideal decompiler is one that transforms all inputs into source code that faithfully reflects the original code: the decompiled code 1) can be recompiled with a Java compiler and 2) behaves the same as the original program. 
However, previous studies that compared Java decompilers \cite{Hamilton2009,Kostelansky2017} found that this ideal Java decompiler does not exist, because of the irreversible data loss that happens during compilation. In this paper, we perform a comprehensive assessment of three aspects of decompilation: the syntactic correctness of the decompiled code (the decompiled code can recompile); the semantic equivalence with the original source (the decompiled code passes all tests); the syntactic similarity to the original source (the decompiled source looks like the original). 
We evaluate eight recent and notable decompilers on $2041$ Java classes, making this study one order of magnitude larger than the related work \cite{Hamilton2009,Kostelansky2017}.

Next, we isolate a subset of $157$ Java classes that no state-of-the-art decompiler can correctly handle. The presence of generics and wildcards is a major challenge that prevents successful decompilation. Meanwhile, we note that different decompilers fail for diverse reasons. This raises the opportunity to merge the results of several incorrect decompiled sources to produce a version that can be recompiled. We call this process \textbf{meta-decompilation}. Meta-decompilation is a novel approach for decompilation: 1) it leverages the natural diversity of existing decompilers by merging the results of different decompilers 2) it is able to provide decompiled sources for classes that no decompiler in isolation can handle. We implement this approach in a novel meta-decompiler called \metadc.

Our results have important implications: 
1) for all users of decompilation, our paper shows significant differences between decompilers and provide well-founded empirical evidence to choose the best ones;
2) for researchers in decompilation, our results show that the problem is not solved;
3) for authors of decompilers, our experiments have identified bugs in their decompilers ($3$ have already been fixed, and counting) and our methodology of \semi can be embedded in the QA process of all decompilers in the world.

In summary, this paper makes the following contributions:
\begin{itemize}
    \item an empirical comparison of eight Java decompilers based on $2041$ real-world Java classes, tested by $25019$ test cases, identifying the key strengths and limitations of Java bytecode decompilation; 
    \item meta-decompilation, a novel approach to decompilation that leverages decompilers diversity to improve decompilation effectiveness;
    \item a tool and a dataset for future research on Java decompilers publicly available at \url{https://github.com/castor-software/decompilercmp}
    
\end{itemize}

\section{Background}\label{sec:background}

\sloppy
In this section, we present an example drawn from the Apache \texttt{commons-codec} library. We wish to illustrate information loss during compilation of Java source code, as well as the different strategies that bytecode decompilers adopt to cope with this loss when they generate source code from bytecode. \autoref{lst:commons-codec-util-original-src} shows the original source code of the utility class \texttt{org.apache.commons.codec.net.Utils}, while \autoref{lst:commons-codec-util-bytecode} shows an excerpt of the bytecode produced by the standard \javac compiler.\footnote{There are various Java compilers available, notably Oracle \javac and Eclipse \ecj, which can produce different bytecode for the same Java input.} Here, we omit the constant pool as well as the table of local variables and replace references towards these tables with comments to save space and make the bytecode more human readable. 

As mentioned, the key challenge of decompilation resides in the many ways in which information is lost during compilation. Consequently, Java decompilers need to make several assumptions when interpreting bytecode instructions, which can also be generated in different ways. To illustrate this phenomenon, \autoref{lst:commons-codec-util-fernflower-src} and \autoref{lst:commons-codec-util-dava-src} show the Java sources produced by the \fernflower and \dava decompilers when interpreting the bytecode of \autoref{lst:commons-codec-util-bytecode}. In both cases, the decompilation produces correct Java code (\ie, recompilable) with the same functionality as the input bytecode. Notice that \fernflower guesses that the series of \texttt{StringBuilder} (bytecode instruction $23$ to $27$) calls is the compiler's way of translating string concatenation and is able to revert it. On the contrary, the \dava decompiler does not reverse this transformation. 

As we notice, the decompiled sources are different from the original in at least three points: 
1) In the original sources, the local variable $i$ was \texttt{final}, but \javac lost this information during compilation.
2) The \texttt{if} statement had originally no \texttt{else} clause. Indeed, when an exception is thrown in a method that does not catch it, the execution of the method is interrupted. Therefore, leaving the \texttt{return} statement outside of the \texttt{if} is equivalent to putting it inside an \texttt{else} clause.
3) In the original code the String \texttt{"Invalid URL encoding: not a valid digit (radix 16): "} was actually computed with \texttt{"Invalid URL encoding: not a valid digit (radix " + URLCodec.RADIX + "): "}. In this case, \texttt{URLCodec.RADIX} is actually a final static field that always contains the value $16$ and cannot be changed. Thus, it is safe for the compiler to perform this optimization, but the information is lost in the bytecode.

\begin{lstlisting}[style=Java, float, floatplacement=H, caption={Source code of Java class correspondig to \texttt{org.apache.commons.codec.net.Utils}.}, label={lst:commons-codec-util-original-src},belowskip=-6\baselineskip] 
class Utils {
    private static final int RADIX = 16;
    static int digit16(final byte b) throws DecoderException {
        final int i = Character.digit((char) b, RADIX);
        if (i == -1) {
            throw new DecoderException("Invalid URL encoding: not a valid digit (radix " + RADIX + "): " + b);
        }
        return i;
    }
}
\end{lstlisting}
\begin{lstlisting}[style=Java, float, floatplacement=H, caption={Excerpt of disassembled bytecode from code in \autoref{lst:commons-codec-util-original-src}.}, label={lst:commons-codec-util-bytecode},belowskip=-6\baselineskip] 
class org.apache.commons.codec.net.Utils {
  static int digit16(byte) throws org.apache.commons.codec.DecoderException;
         0: ILOAD_0             //Parameter byte b
         1: I2C
         2: BIPUSH        16
         4: INVOKESTATIC  #19   //Character.digit:(CI)I            
         7: ISTORE_1            //Variable int i
         8: ILOAD_1
         9: ICONST_m1
        10: IF_ICMPNE     37
//org/apache/commons/codec/DecoderException
        13: NEW           #17
        16: DUP
        17: NEW           #25   //java/lang/StringBuilder
        20: DUP
//"Invalid URL encoding: not a valid digit (radix 16):"
        21: LDC           #27
//StringBuilder."<init>":(Ljava/lang/String;)V
        23: INVOKESPECIAL #29
        26: ILOAD_0
//StringBuilder.append:(I)Ljava/lang/StringBuilder;
        27: INVOKEVIRTUAL #32
//StringBuilder.toString:()Ljava/lang/String;
        30: INVOKEVIRTUAL #36
//DecoderException."<init>":(Ljava/lang/String;)V
        33: INVOKESPECIAL #40
        36: ATHROW
        37: ILOAD_1
        38: IRETURN
}
\end{lstlisting}

\begin{lstlisting}[style=Java, float, floatplacement=H, caption={Decompilation result of \autoref{lst:commons-codec-util-bytecode} with \fernflower.}, label={lst:commons-codec-util-fernflower-src},belowskip=-6\baselineskip] 
class Utils {
   private static final int RADIX = 16;
   static int digit16(byte b) throws DecoderException {
      int i = Character.digit((char)b, 16);
      if(i == -1) {
         throw new DecoderException("Invalid URL encoding: not a valid digit (radix 16): " + b);
      } else {
         return i;
      }
   }
}
\end{lstlisting}
\begin{lstlisting}[style=Java, float, floatplacement=H, caption={Decompilation result of \autoref{lst:commons-codec-util-bytecode} with \dava.}, label={lst:commons-codec-util-dava-src},belowskip=-6\baselineskip] 
class Utils
{
    static int digit16(byte b)
        throws DecoderException
    {
        int i = Character.digit((char)b, 16);
        if(i == -1)
            throw new DecoderException((new StringBuilder()).append("Invalid URL encoding: not a valid digit (radix 16): ").append(b).toString());
        else
            return i;
    }
    private static final int RADIX = 16;
}
\end{lstlisting}

Besides, this does not include the different formatting choices made by the decompilers such as new lines placement and brackets usage for single instructions such as \texttt{if} and \texttt{else}.

\section{Decompiler evaluation methodology}\label{sec:methodology}

In this section, we introduce definitions, metrics and research questions. Next, we  detail the framework to compare decompilers and we describe the Java projects that form the set of case studies for this work.

\subsection{Definitions and Metrics}
\label{sec:def}

The value of the results produced by decompilation varies greatly depending on the intended use of the generated source code. In this work, we evaluate the decompilers' capacity to produce a faithful retranscription of the original sources. Therefore, we collect the following metrics.

\begin{defn}
\label{met:syntactic-c}
\textbf{Syntactic correctness.} The output of a decompiler is syntactically correct if it contains a valid Java program, \ie a Java program that is recompilable with a Java compiler without any error.
\end{defn}

When a bytecode decompiler generates source code that can be recompiled, this source code can still be  syntactically different from the original. 
We introduce a metric to measure the scale of such a difference according to the abstract syntax tree (AST) dissimilarity\cite{Falleri2014} between the original and the decompiled results. This metric, called \textit{syntactic distortion}, allows to measure the differences that go beyond variable names. The description of the metric is as follows:

\begin{defn}
\label{def:syntactic-distortion}
\textbf{Syntactic distortion.} The minimum number of atomic edits required to transform the AST of the original source code of a program into the AST of the corresponding decompiled version of it. 
\end{defn}

In the general case, determining if two program are semantically equivalent is undecidable. For some cases, the decompiled sources can be recompiled into bytecode that is equivalent to the original, modulo reordering of the constant pool. We call these cases \textit{strictly equivalent} programs. We measure this equivalence with a bytecode comparison tool named Jardiff.\footnote{\url{https://github.com/scala/jardiff}}

Inspired by the work of~\cite{Le2014} and~\cite{Yanghunting2019}, we check if the decompiled and recompiled program are semantically equivalent modulo inputs. This means that for a given set of inputs, the two programs produce equivalent outputs. In our case, we select the set of relevant inputs and assess equivalence based on the existing test suite of the original program.

\begin{defn}
\label{met:semi}
\textbf{Semantic equivalence modulo inputs.} We call a decompiled program semantically equivalent modulo inputs to the original if it passes the set of tests from the original test suite.
\end{defn}

In the case where the decompiled and recompiled program produce non-equivalent outputs, that demonstrates that the sources generated by the decompiler express a different behavior than the original. As explained by Hamilton and colleagues \cite{Hamilton2009}, this is particularly problematic as it can mislead decompiler users in their attempt to understand the original behavior of the program. We refer to theses cases as \textit{deceptive decompilation} results.

\begin{defn}
\label{def:deceptive}
\textbf{Deceptive decompilation:} Decompiler output that is syntactically correct but not semantically equivalent to the original input.
\end{defn}

\subsection{Research Questions}

We elaborated five research questions to guide our study on the characteristics of modern Java decompilers. 

\newcommand{\RQone}{To what extent is decompiled Java code syntactically correct?}
\newcommand{\RQtwo}{To what extent is decompiled Java code semantically equivalent modulo inputs?}
\newcommand{\RQthree}{To what extent do decompilers produce deceptive decompilation results?}
\newcommand{\RQfour}{What is the syntactic distortion of decompiled code?}
\newcommand{\RQfive}{To what extent do the successes and failures of decompilers overlap?}

\textbf{RQ1: \RQone} In this research question, we investigate the effectiveness of decompilers for producing syntactically correct and hence recompilable source code from bytecode produced by the \javac and \ecj compilers.

\textbf{RQ2: \RQtwo}  Le and colleagues \cite{Le2014} propose to use equivalence modulo inputs assessment as a way to test transformations that are meant to be semantic preserving (in particular compilation). In this research question, we adapt this concept in the context of decompilation testing. In this paper we rely on the existing test suite instead of generating inputs.

\textbf{RQ3: \RQthree}
In this research question, we investigate the cases where we observe semantic differences between the original source code and the outputs of the decompilers.

\textbf{RQ4: \RQfour} Even if decompiled bytecode is ensured to be syntactically and semantically correct, syntactic differences may remain as an issue when the purpose of decompilation is human understanding. Keeping the decompiled source code free of syntactic distortions is essential during program comprehension, as many decompilers can produce human unreadable code structures. In this research question, we compare the syntactic distortions produced by decompilers.

\textbf{RQ5: \RQfive} In this research question we investigate the intersection of classes for which each decompiler produce semantically equivalent modulo input sources.

\subsection{Study Protocol}

\begin{figure}
	\centering
	\includegraphics[origin=c,width=0.48\textwidth]{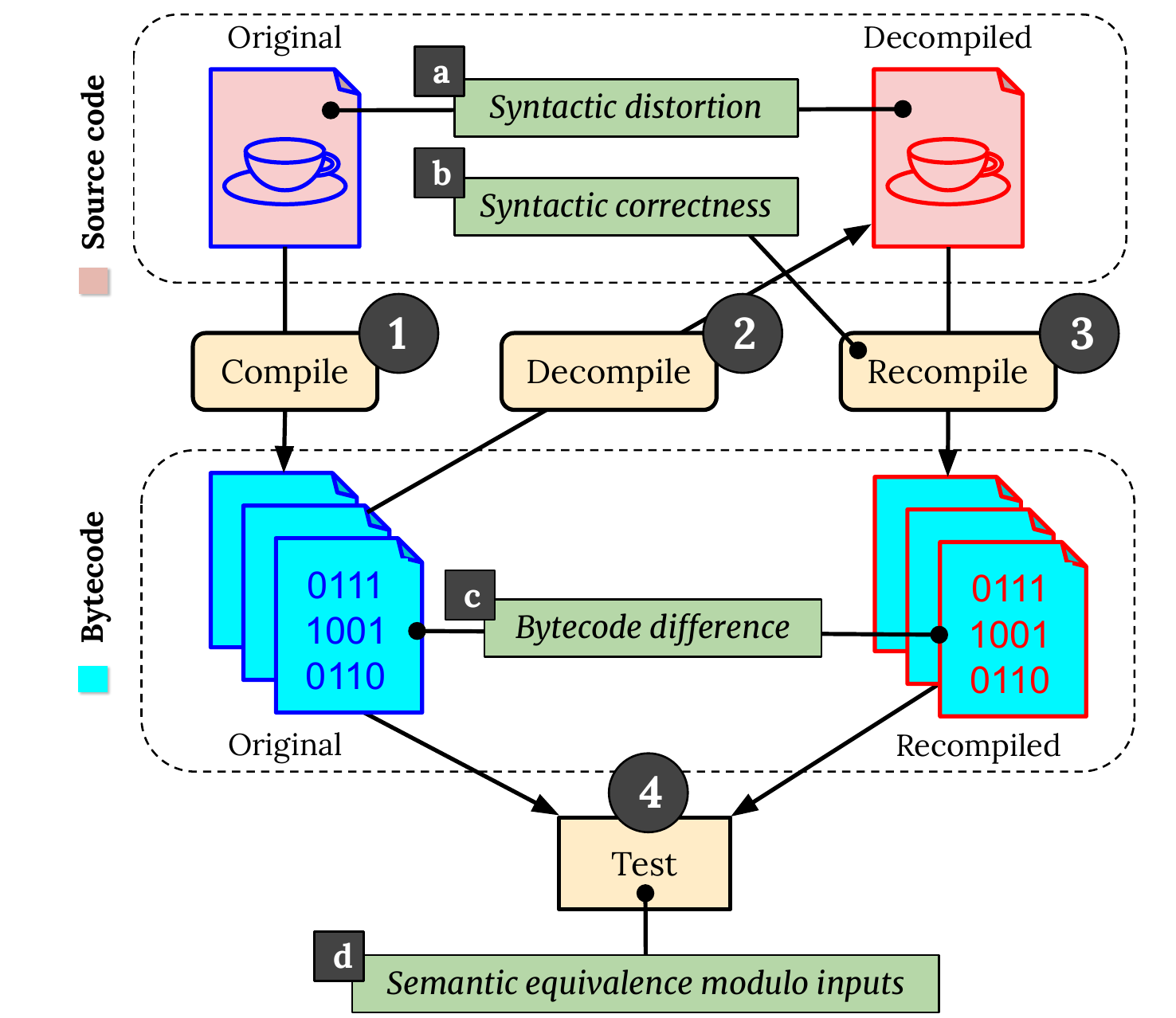}
	\caption{Java decompiler assessment pipeline with four evaluation layers: syntactic distortion, bytecode difference, syntactic correctness, and semantic equivalence modulo input.}
	\label{fig:pipeline}
\end{figure}

\autoref{fig:pipeline} represents the pipeline of operations conducted on every Java source file in our dataset. For each triplet \textit{<decompiler, compiler, project>}, we perform the following: 
\begin{enumerate}
    \item Compile the source files with a given compiler.
    \item Decompile each class file  with a decompiler (there might be several classes if the source defines internal classes). If the decompiler does not return any error, we mark the source file as decompilable. Then, (a) we measure syntactic distortion by comparing the AST of the original source with the AST of the decompiled source.
    \item Recompile the class files with the given compiler. If the compilation is successful, we know that the decompiler produces (b) syntactically correct code. Then, we measure (c) the textual difference between the original  and the recompiled bytecode. If there are none, the decompiler produced semantically equivalent code, otherwise we cannot assess anything in that regard yet.
    \item Run the test cases on the recompiled bytecode. If the tests are successful, we mark the source as \textit{passTests} for the given triplet, showing that the decompiler produces   (d) semantically equivalent code modulo inputs.
\end{enumerate}

If one of these steps fails we do not perform the following steps and consider all the resulting metrics not available. As decompilation can generate a program that does not stop, we set a $20$ minutes  timeout  for the execution of the test suite (the original test suites run under a minute on the hardware used for this experiment, a Core i5-6600K with 16GB of RAM).

The tests used to assess the semantic equivalence modulo inputs are those of the original project that cover the given Java file.\footnote{Coverage was assessed using yajta \url{https://github.com/castor-software/yajta}} We manually excluded the tests that fail on the original project (either flaky or because versioning issue). The list of excluded tests is available as part of our experiments.

\subsection{Study Subjects}

\textbf{Decompilers.}
\autoref{tab:decompilers} shows the set of decompilers under study. We have selected Java decompilers that are (i) freely available, and (ii) have been active in the last two years. We add \jode in order to compare our results with a legacy decompiler, and because the previous survey by Hamilton and colleagues considers it to be  one of the best decompilers \cite{Hamilton2009}.

The column \textsc{Version} shows the version used (some decompilers do not follow any versioning scheme). We choose the latest release if one exists, if not the last commit available the 09-05-2019. The column \textsc{Status} indicates the date of the last commit or "Active" if the last commit was more recent than 30 days. The column \textsc{\#Commits} represents the number of commits in the decompiler project, in cases where the decompiler is a submodule of a bigger project (\eg \dava and \fernflower) we count only commits affecting the submodule. The column \textsc{\#LOC} is the number of lines of code in all Java files (and Python files for \krakatau) of the decompiler, including sources, test sources and resources counted with \textit{cloc}.\footnote{\url{http://cloc.sourceforge.net/}}

\begin{table}
\caption{Characteristics of the studied decompilers.}
\centering
\scriptsize
\begin{tabular}{lcccc}
			\hline
			\textsc{\textbf{Decompiler}}  & \textsc{\textbf{Version}} & \textsc{\textbf{Status}} & \textsc{\textbf{\#Commits}} & \textsc{\textbf{\#LOC}} \\
			\hline
    		\cfr \cite{cfr}   & $0.141$ & Active & $1433$ & $52098$ \\
    		\dava \cite{dava}        & $3.3.0$ & 2018-06-15* & $14$ & $22884$ \\
    		\fernflower \cite{fernflower}   & NA** & Active & $453$ & $52118$ \\
    		\jadx \cite{jadx} & 0.9.0 & Active & $970$ & $55335$ \\
    		\jd \cite{jdcore} & 1.0.0 & Active & NA*** & $36730$ \\
    		\jode \cite{jode} & 1.1.2-pre1 & 2004-02-25* & NA*** & $30161$ \\
    		\krakatau \cite{krakatau}  & NA** & 2018-05-13* & $512$ & $11301$ \\
    		\procyon \cite{procyon}  & $0.5.34$ & Active & $1080$ & $122147$ \\
            \hline
		\end{tabular}
	
	\label{tab:decompilers}
	\begin{tablenotes}
      \footnotesize
      \centering
      \item * Date of last update.
      \item ** Not following any versioning scheme.
      \item *** CVS not available at the date of the present study.
    \end{tablenotes}
    \vspace{-0.5cm}
\end{table}

Note that different decompilers are developed for different usages and, therefore, are meant to achieve different goals.
\cfr \cite{cfr} for Java 1\footnote{\url{https://github.com/leibnitz27/cfr/blob/33216277ae3b61a9d2b3f912d9ed91a3e698d536/src/org/benf/cfr/reader/entities/attributes/AttributeCode.java\#L49}} to 14, for code compiled with \javac (note that since we performed our first experiments, it is now tested with \ecj generated classes).
\procyon \cite{procyon} from Java 5 and beyond and \javac, shares its test suite with \cfr.
\fernflower \cite{fernflower} is the decompiler embedded in IntelliJ IDE.
\krakatau \cite{krakatau} up to Java 7 does not currently support Java 8 or \texttt{invokedynamic}.
\jd \cite{jdcore} is the engine of JD-GUI. It supports Java 1.1.8 to Java 12.0. The version we study in this work is the first version released since the complete rewrite of JD-Core. While older versions were based on a simple bytecode pattern recognition engine, \jd now includes a CFG analysis layer.
\jadx \cite{jadx} is a decompiler that originally targeted dex files (bytecode targeting the android platform) but can also target class files, as in our experiments.
\dava \cite{dava} is a decompiler built on top of the Soot Framework \cite{soot}. It does not target Java bytecode produced by any specific compiler nor from any specific language, but produces decompiled sources in Java. Soot supports bytecode and source code up to Java 7.
\jode \cite{jode} is a legacy decompiler that handles Java bytecode up to Java 1.4.

\textbf{Projects.}
In order to get a set of real world Java projects to evaluate the eight decompilers, we reuse the set of projects of Pawlak and colleagues\cite{Pawlak2015}. To these $13$ projects we added a fourteenth one named \texttt{DcTest} made out of examples collected from previous decompiler evaluations \cite{Hamilton2009,Kostelansky2017}.\footnote{\url{http://www.program-transformation.org/Transform/JavaDecompilerTests}} \autoref{tab:dataset} shows a summary of this dataset: the Java version in which they are written, the number of Java source files, the number of unit tests as reported by Apache Maven, and the number of Java lines of code in their sources.

As different Java compilers may translate the same sources into different bytecode representations \cite{sootdiff}\footnote{\url{https://www.benf.org/other/cfr/eclipse-differences.html}}, we experiment with the two most used Java compilers: \javac and \ecj (versions $1.8.0\_17$ and $13.13.100$, respectively). We compiled all $14$ projects with both compilers (except \texttt{commons-lang} which failed to build with \ecj). 
Our dataset includes $3928$ bytecode classes, $1887$ of which compiled with \ecj, and $2041$ compiled with \javac.
As we study the influence of the compiler, in RQ1, we limit our datasets to the $1887$ classes that compiled with both compilers.
As semantic equivalence modulo inputs is based on test suites, for RQ2 and RQ3 we focus on the classes that contain code executed by test suites: $2397$ classes generated by the two compilers. These classes covered by the test suites exclude interfaces as they do not contain executable code. Most enum declarations fall under the same category. Test coverage is assessed through bytecode instrumentation with a tool named yajta.\footnote{\url{https://github.com/castor-software/yajta}}

\begin{table}
\caption{Characteristics of the projects used to evaluate decompilers.}
\scriptsize
\centering
    \begin{tabular}{lcccc}
		\hline
		\textsc{\textbf{Project name}} & \textsc{\textbf{Java version}} & \textsc{\textbf{\#Classes}} & \textsc{\textbf{\#Tests}} & \textsc{\textbf{\#LOC}}\\
		\hline
        Bukkit	& $1.6$ & $642$ & $906$ & $60800$\\
        Commons-codec & $1.6$ & $59$ & $644$ & $15087$\\
        Commons-collections	& $1.5$ & $301$ & $15067$ & $62077$\\
        Commons-imaging	& $1.5$ & $329$ & $94$ & $47396$\\
        Commons-lang & $1.8$ & $154$ & $2581$ & $79509$\\
        DiskLruCache & $1.5$ & $3$ & $61$ & $1206$\\
        JavaPoet*	& $1.6$ & $2$ & $60$ & $934$\\
        Joda time	& $1.5$ & $165$ & $4133$ & $70027$\\
        Jsoup	& $1.5$ & $54$ & $430$ & $14801$\\
        JUnit4	& $1.5$ & $195$ & $867$ & $17167$\\
        Mimecraft	& $1.6$ & $4$ & $14$ & $523$\\
        Scribe Java	& $1.5$ & $89$ & $99$ & $4294$\\
        Spark & $1.8$ & $34$ & $54$ & $4089$\\
        DcTest** & $1.5-1.8$ & $10$ & $9$ & $211$\\
        \hline
        \textsc{\textbf{Total}} &  & \textbf{$2041$}  & \textbf{$25019$} & \textbf{$378121$}\\
        \hline
	\end{tabular}

	\label{tab:dataset}
	\begin{tablenotes}
	\centering
      \footnotesize
      \item (*) Formerly named JavaWriter.
      \item (**) Examples collected from previous decompilers evaluation.
    \end{tablenotes}
    \vspace{-0.5cm}
\end{table}

\section{Experimental Results}\label{sec:results}

\subsection{\textbf{RQ1: (syntactic correctness)} \RQone}
\label{sec:rq1-results}
This research question  investigates to what extent the source code produced by the different decompilers is syntactically correct, meaning that the decompiled code compiles. We also investigate the effect of the compiler that produces the bytecode on the decompilation results. To do so, in this section, we focus on the $1887$ classes that compile with both \javac and \ecj.

\begin{figure}
	\centering
	\includegraphics[origin=c,width=0.485\textwidth]{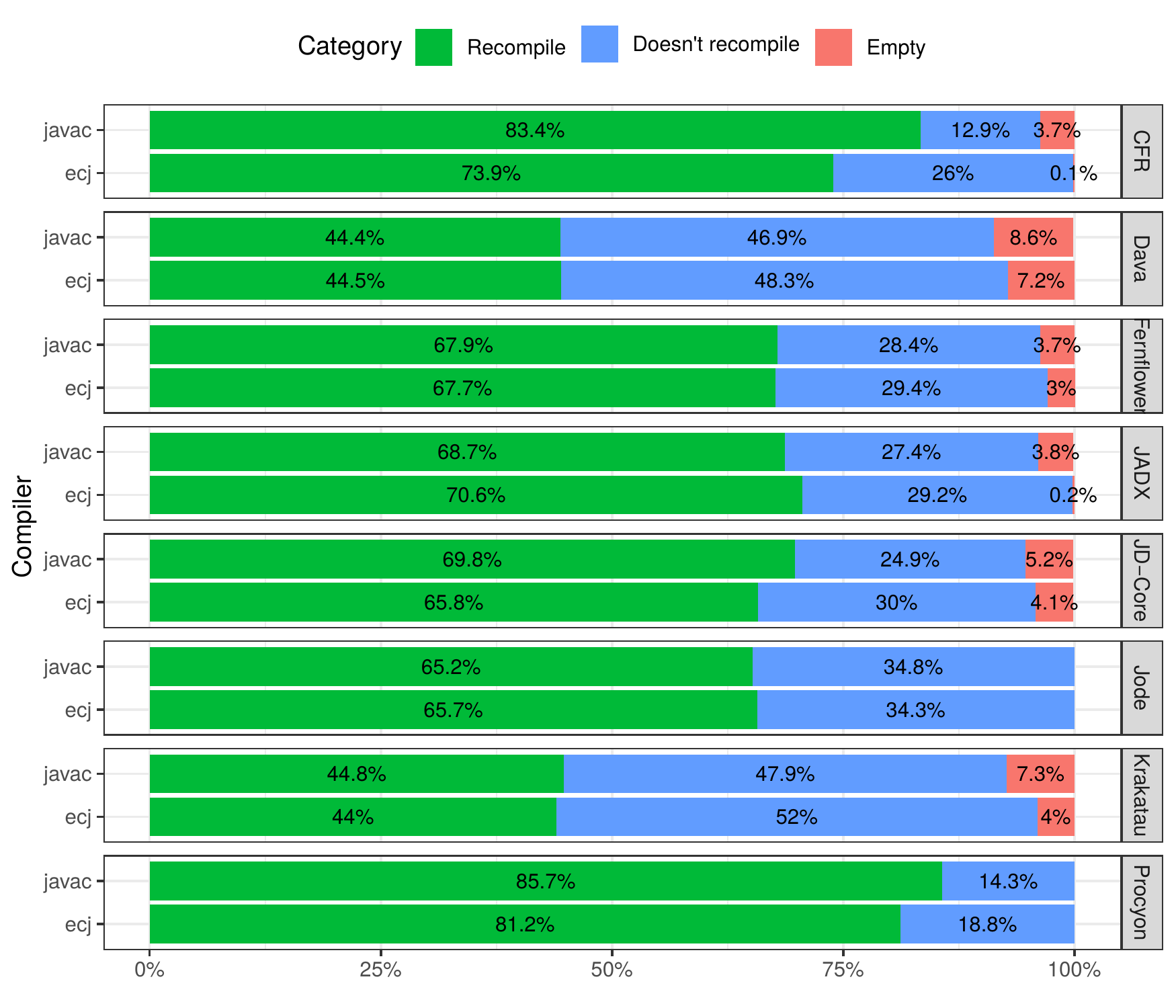}
	\caption{Outcome of the decompilation for each pair of compiler and decompiler for the $1887$ classes compilable by both \ecj and \javac. From left to right are presented the percentages of classes that were syntactically correct (in green), syntactically incorrect (in blue), and empty (in red)}
	\label{fig:decompilation_categories_results}
	\vspace{-0.6cm}
\end{figure}

\autoref{fig:decompilation_categories_results} shows the ratio of decompiled classes that are syntactically correct per pair of compiler and decompiler. The horizontal axis shows the ratio of syntactically correct output in green, the ratio of syntactically incorrect output in blue, and the ratio of empty output in red (an empty output occurs, e.g. when the decompiler crashes). The vertical axis shows the compiler on the left and decompiler on the right. For example, \procyon, shown in the last row, is able to produce syntactically correct source code for $1609$ ($85.3\%$) class files compiled with \javac, and produce a non-empty syntactically incorrect output for $278$ ($14.7\%$) of them. On the other hand, when sources are compiled with \ecj, \procyon generates syntactically correct sources for $1532$ ($81.2\%$) of class files and syntactically incorrect for $355$ ($18.8\%$) sources. In other words, \procyon is slightly more effective when used against code compiled with \javac. 
It is interesting to notice that not all decompiler authors have decided to handle error the same way. Both \procyon and \jode's developers have decided to always return source files, even if incomplete (for our dataset). Additionally, when \cfr and \procyon detect a method that they cannot decompile properly, they may replace the body of the method  by a single \texttt{throw} statement and comment explaining the error. This leads to syntactically correct code, but not semantically equivalent.

The ratio of syntactically correct decompiled code ranges from $85.7\%$ for \procyon on \javac inputs (the best), down to $44\%$ for \krakatau on \ecj (the worst). All decompilers failed to produce syntactically correct output for 137 classes.
Overall, no decompiler is capable of correctly handling the complete dataset. This illustrates the challenges of Java bytecode decompilation, even for bytecode that has not been obfuscated.

We note that syntactically incorrect decompilation can still be useful for reverse engineering. However, an empty output is useless: the ratio of class files for which the decompilation completely fails is never higher than $8.6\%$ for \dava on \javac bytecode.

\textbf{}\begin{lstlisting}[style=Java, float, floatplacement=H, caption={Excerpt of differences in \texttt{Bag} original (in red marked with a -) and decompiled with \cfr  (in green marked with a +).}, label={lst:bag-original-cfr},belowskip=-6\baselineskip]
public interface Bag<E> extends Collection<E> {
£+£¤aa¤£@Override£
£+£¤aa¤£public boolean add(E var1);£
`-`µaaµ`public boolean add(E Object);`
    [...]
}
\end{lstlisting}
In the following paragraphs we investigate the impact of the compiler on decompilation effectiveness. Over the 10912 syntactically incorrect decompilation (over all decompilers), 712 can be attributed to compiler differences because the decompiler produces syntactically incorrect sources for one compiler and not for the other. These cases break down as follows: 596 failures occur only on  \ecj bytecode and 116 cases only on \javac code.

\autoref{lst:bag-original-cfr} shows an excerpt of the differences between the original source code of the  \texttt{Bag} interface from  \texttt{commons-collections} and its decompiled sources produced by \cfr. This is an example where both \javac and \ecj produce the same bytecode, yet recompilation of the sources produced by \cfr succeed with \javac and fail with \ecj. The \texttt{commons-collections} library is compiled targeting Java 1.5, in our experiment. In these conditions, \ecj fails in the presence of  the  \texttt{@Override} annotation to override a method inherited from an interface such as \texttt{Collection}. It is only accepted for inheritance from a class (abstract or concrete). Meanwhile, \javac compiles the decompiled sources without any error. Note that  specifying Java 1.6  as target solves the error with \ecj. This illustrates how the notion of syntactic correctness depends on the actual compiler as well as on the targeted Java version.

\textbf{}\begin{lstlisting}[style=Java, float, floatplacement=H, caption={Excerpt of differences when compiling  \texttt{org/apache/\allowbreak{}commons/\allowbreak{}codec/\allowbreak{}language/\allowbreak{}bm/\allowbreak{}Lang} with \javac (in red marked with a -) and with \ecj  (in green marked with a +). Top part of listing illustrates differences in bytecode, while bottom part shows differences in source code decompiled by \procyon.}, label={lst:lang-original-procyon},belowskip=-6\baselineskip]
//Bytecode
   NEW           Lang$LangRule
   DUP
   ALOAD         10 //pattern
   NEW           HashSet
   DUP
   ALOAD         11 //langs
   INVOKESTATIC  Arrays.asList
   invokespecial HashSet.<init>
   ILOAD         12 // accept
   ACONST_NULL
`-`µaaµ`INVOKESPECIAL Lang$LangRule.<init>`
`-`µaaaaµ`(LPattern;LSet;Z LLang$1)V`
£+£¤aa¤£INVOKESPECIAL Lang$LangRule.<init>£
£+£¤aaaa¤£(LPattern;LSet;Z LLang$LangRule)V£

//Decompiled sources
//Usage of private static inner class LangRule constructor's wrapper generated by the compiler
`-`µaaµ`new LangRule(pattern, new HashSet<String>(`
`-`µaaaaaaaaµ`Arrays.asList(langs)), accept)`
£+£¤aa¤£new LangRule(pattern, new HashSet(£
£+£¤aaaaaaaa¤£Arrays.asList(langs)), accept, null)£
\end{lstlisting}
\autoref{lst:lang-original-procyon} shows an excerpt of the bytecode generated by \javac and \ecj for  class \texttt{org/apache/\allowbreak{}commons/\allowbreak{}codec/\allowbreak{}language/\allowbreak{}bm/\allowbreak{}Lang} as well as the corresponding decompiled sources generated by \procyon in both cases. The excerpt shows a call to the \texttt{private} constructor of a \texttt{private static} nested class of \texttt{Lang} called \texttt{LangRule}. Since the nested class is \texttt{static} the outer class and the inner class interact as if both were top-level classes\footnote{https://docs.oracle.com/javase/tutorial/java/javaOO/nested.html} in the bytecode. But as the constructor of the nested class is \texttt{private}, the enclosing class cannot access it. To bypass this problem, both \javac and \ecj create a synthetic \texttt{public} wrapper for this constructor. In Java bytecode, a synthetic element is an element created by the compiler that does not correspond to any element present in the original sources (implicitly or not). This wrapper is a public constructor for the nested class \texttt{LangRule}. As the signature for this wrapper cannot be the same as the private constructor, it must have different parameters. \javac and \ecj handle this case differently. \javac creates a synthetic anonymous class \texttt{Lang\$1} and adds  an additional parameter typed with this anonymous class to the wrapper parameters. Since the value of this parameter is never used, null is passed as additional parameter when the wrapper is called. \ecj does almost the same thing, but the additional parameter is of type \texttt{Lang\$LangRule}. \procyon is able to reverse  the \javac transformation  correctly, but not the \ecj one. It decompiles the \ecj version literally,  conserving the null parameter. Yet,  the synthetic wrapper  does not exist in the decompiled sources. Consequently, the decompiled code that refers to an absent constructor  is syntactically incorrect. 
This example illustrates the decompilation challenge introduced by synthetic elements that are generated by a compiler. Note that synthetic elements in bytecode do carry a flag indicating their nature. But, (i) it does not change the difficulty of reversing an unforeseen pattern and (ii) this flag could be abused by an obfuscator as it does not change the semantic of the bytecode but rather gives indication to drive potential tools modifying the bytecode. In the case of \procyon and \cfr, their common test suite did not include code generated by \ecj at the time of this experiment. Since then, the author of \cfr has updated the test suite common to \cfr and \procyon with test covering bytecode from \ecj\footnote{Commit: \url{https://github.com/leibnitz27/cfr_tests/commit/b4f0b01e34a953a1fd57e52f508c9c02c58e6dee} Discussion: \url{https://github.com/leibnitz27/cfr/issues/50}}

We have shown that the compiler impact decompilation effectiveness for two reasons: (i) it changes the oracle for syntactic correctness; (ii) it produces different bytecode structure that decompilers might not expect.\looseness=-1

To assess the significance of this impact, we use a $\chi^{2}$ test on the ratio of classfiles decompiled into syntactically correct source code depending on the compiler, \javac versus \ecj.
The compiler variable has an impact for three decompilers and no impact for the remaining five, with $99\%$ confidence level.
The test rejects that the compiler has no impact on the decompilation syntactic correctness ratio for  \cfr, \procyon and \jd (p-value $10^{-14}$, $0.00027$ and $0.006444$). For the five other decompilers we do not observe a significant difference between \javac and \ecj (p-values: \dava $0.15$, \fernflower $0.47$, \jadx $0.17$, \jode $0.50$, and \krakatau $0.09$). Note that beyond syntactic correctness, the compiler may impact the correctness of the decompiled code, this is discussed in more details in Section~\ref{sec:rq3-results}.

To sum up, \procyon and \cfr are the decompilers that score the highest on syntactic correctness. The three decompilers ranking the lowest are \jode, \krakatau and \dava. It is interesting to note that those three are no longer actively maintained.\looseness=-1
\begin{mdframed}[style=mpdframe]
\textbf{Answer to RQ1:} 
No single decompiler is able to produce syntactically correct sources for more than $85.7\%$ of class files in our dataset. The implication for decompiler users is that decompilation of Java bytecode cannot be blindly applied and does require some additional manual effort. Only few cases make all decompilers fail, which suggests that using several decompilers in conjunction could help to achieve better results.
\end{mdframed}

\subsection{\textbf{RQ2: (semantic equivalence)} \RQtwo}

\begin{lstlisting}[style=Java, float, floatplacement=H, caption={Excerpt of bytecode from class \texttt{org/apache/\allowbreak{}commons/\allowbreak{}codec/\allowbreak{}language/\allowbreak{}bm/\allowbreak{}Lang.class}, compiled with \javac and decompiled with \cfr:
    Lines in red, marked with a -, are in the original bytecode, while lines in green, marked with a +, are from the recompiled sources.}, label={lst:lang-bytecode},belowskip=-6\baselineskip]
`-`µaaaµ`IFEQ L2`
£+£¤aaa¤£IFNE L2£
£+£¤aaa¤£GOTO L0£
£+ L2£
     ALOAD 5
     INVOKESTATIC Lang$LangRule.access$100 (LLang$LangRule;)Z
     IFEQ L3
     ALOAD 3
     ALOAD 5
     INVOKESTATIC Lang$LangRule.access$200 (Lang$LangRule;)Ljava/util/Set;
`-`µaaaµ`INVOKEINTERFACE Set.retainAll (LCollection;)Z` 
£+£¤aaa¤£INVOKEVIRTUAL HashSet.retainAll (LCollection;)Z£
     (itf)
     POP
`-`µaaaµ`GOTO L2`
£+£¤aaa¤£GOTO L0£
\end{lstlisting}

To answer this research question, we  focus on the $2397$ class files, regrouping bytecode generated by both \javac and \ecj, that are covered by at least one test case. This excludes all types which contain no executable code, such as interfaces. When decompilers produce sources that compile, we investigate the semantic equivalence of the decompiled source and their original. To do so, we split recompilable outputs in three categories: (i) \textit{semantically equivalent:} the code is recompiled into bytecode that is strictly identical to the original (modulo reordering of the constant pool, as explained in Section \ref{sec:def}), (ii) \textit{semantically equivalent modulo inputs:} the output is recompilable and passes the original project's test suite (i.e. we cannot prove that the decompiled code is semantically different), and (iii) \textit{semantically different:} the output is recompilable but it does not pass the original test suite (deceptive decompilation, as explained in Definition \ref{def:deceptive}).

Let us first discuss an example of semantic equivalence of decompiled code. \autoref{lst:lang-bytecode} shows an example of bytecode that is different when decompiled-recompiled but equivalent modulo inputs to the original. Indeed, we can spot two differences: the control flow blocks are not written in the same order (\texttt{L2} becomes \texttt{L0}) and the condition evaluated is reversed (\texttt{IFEQ} becomes \texttt{IFNEQ}), which leads to an equivalent control flow graph. The second difference is that the type of a variable originally typed as a \texttt{Set} and instantiated with an \texttt{HashSet} has been transformed into a variable typed as an \texttt{HashSet}, hence once \texttt{remainAll} is invoked on the variable \texttt{INVOKEINTERFACE} becomes directly \texttt{INVOKEVIRTUAL}. This example illustrates how bytecode may be changed (here with a slightly different but equivalent control flow graph and a change in the precision of the type information) yet still be semantically equivalent modulo inputs.\looseness=-1

\begin{figure}
	\centering
	\includegraphics[origin=c,width=0.47\textwidth]{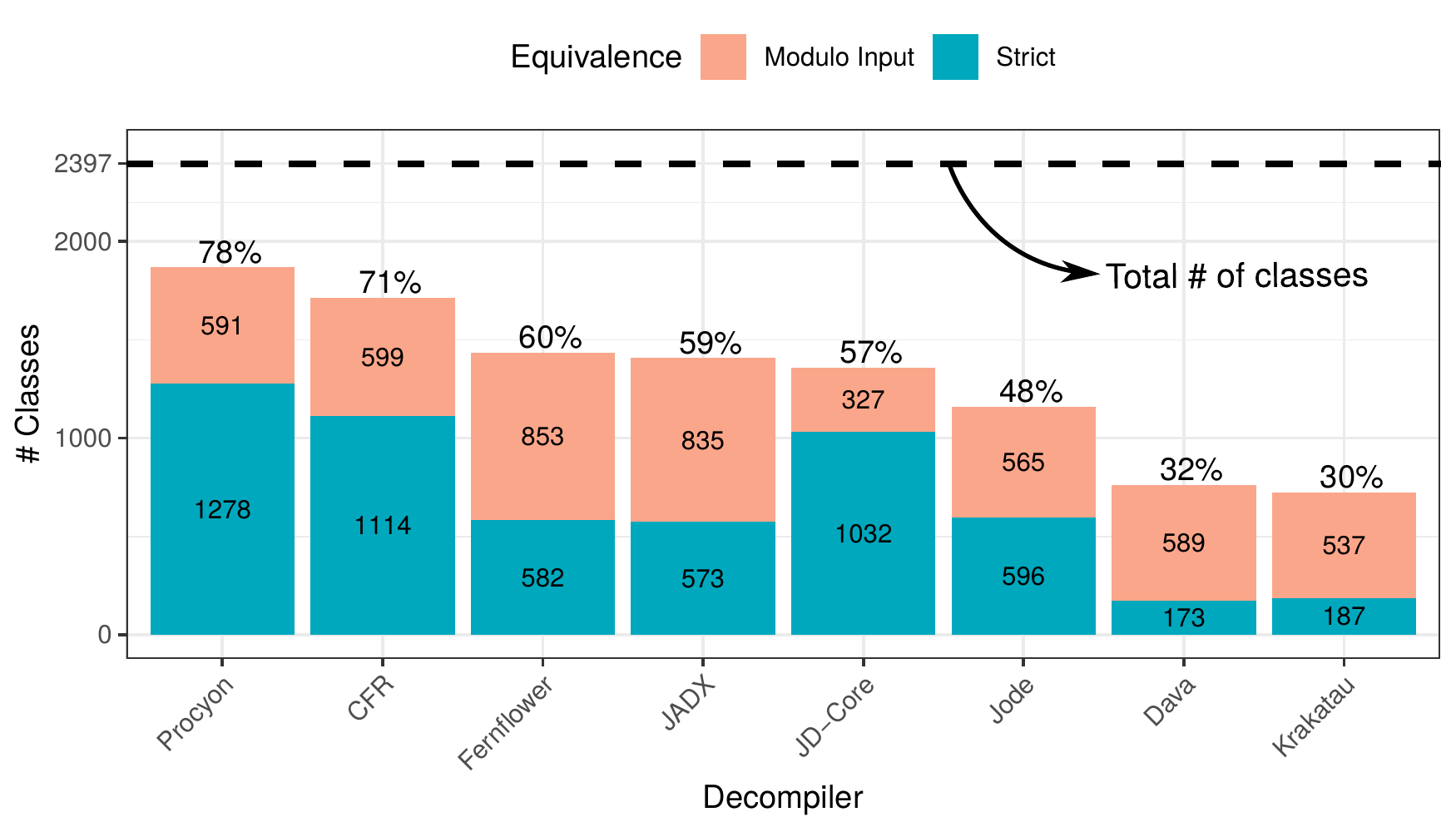}
	\caption{Equivalence results for each decompiler on all the $2397$ classes of the studied projects covered by at least one test, (both compilers combined).}
	\label{fig:equivalence_categories_results}
    \vspace{-0.6cm}
\end{figure}

Now we discuss the results globally. \autoref{fig:equivalence_categories_results} shows the recompilation outcomes of decompilation regarding semantic equivalence for the $2397$ classes under study.
The horizontal axis shows the eight different decompilers. The vertical axis shows the number of classes decompiled successfully.
Strictly equivalent output is shown in blue, equivalent classes modulo input are shown in orange. For example, \cfr (second bar) is able to correctly decompile $1713$ out of $2397$ classes ($71\%$), including $1114$ classes that are recompilable into strictly equivalent bytecode, and $599$ that are recompilable into equivalent bytecode modulo inputs. When the decompiler fails to produce an output that is semantically equivalent modulo  inputs, it is either because the decompiled sources were not syntactically correct and did not recompile, or because they did recompile but did not pass the original test suite (deceptive decompilation). \autoref{tab:failures} gives this information for each decompiler.

\begin{table}
    \centering
    \scriptsize
    \caption{Cause of non-equivalence for each decompiler.}
    \begingroup
    \setlength{\tabcolsep}{5pt} 
    \renewcommand{\arraystretch}{1}
    \begin{tabular}{lcccccccc}
     &
     \rotatebox{90}{\textsc{\textbf{\procyon}}} &
     \rotatebox{90}{\textsc{\textbf{\cfr}}} &
     \rotatebox{90}{\textsc{\textbf{\fernflower}}} &
     \rotatebox{90}{\textsc{\textbf{\jadx}}} &
     \rotatebox{90}{\textsc{\textbf{\jd}}} &
     \rotatebox{90}{\textsc{\textbf{\jode}}} &
     \rotatebox{90}{\textsc{\textbf{\dava}}} &
     \rotatebox{90}{\textsc{\textbf{\krakatau}}}\\ 
      \hline
      \tiny{\#Deceptive}  & 33  & 22  & 21  & 78  & 44   & 142  & 36   & 97   \\
      \tiny{\#!Recompile} & 495 & 662 & 941 & 911 & 994  & 1094 & 1599 & 1576 \\ 
      \hline
      \tiny{\#Failures}   & 528 & 684 & 962 & 989 & 1038 & 1236 & 1635 & 1673 \\ 
      \hline
    \end{tabular}
    \endgroup
    \label{tab:failures}
\end{table}

The three decompilers that are not actively maintained any more (\jode, \dava and \krakatau) handle less than $50\%$ of the cases correctly (recompilable and pass tests). On the other hand, \procyon and \cfr have the highest ratio of equivalence modulo inputs of $78\%$ and $71\%$, respectively. 

\begin{mdframed}[style=mpdframe]
\textbf{Answer to RQ2:} The number of classes for which the decompiler produces  equivalent sources modulo input varies significantly from one decompiler to another. The result of decompilation is usually not strictly identical to the original source code. Five decompilers generate equivalent modulo input source code for more than 50\% of the classes.
For end users, it means that the state of the art of Java decompilation does not guarantee semantically correct decompilation, and care must be taken not to blindly trust the behavior of decompiled code.
\looseness=-1
\end{mdframed}

\subsection{\textbf{RQ3: (bug finding)} \RQthree}

\label{sec:rq3-results}
As explained by Hamilton and colleagues~\cite{Hamilton2009}, while a syntactically incorrect decompilation output may still be useful to the user, syntactically correct but semantically different output is more problematic. Indeed, this may mislead the user by making her believe in a different behaviour than the original program has. We call this case \textit{deceptive decompilation} (as explained in  Definition \ref{def:deceptive}). When such cases occur, since the decompiler produces an output that is semantically different from what is expected, they may be considered decompilation bugs.

\begin{figure}
	\centering
	\includegraphics[origin=c,width=0.475\textwidth]{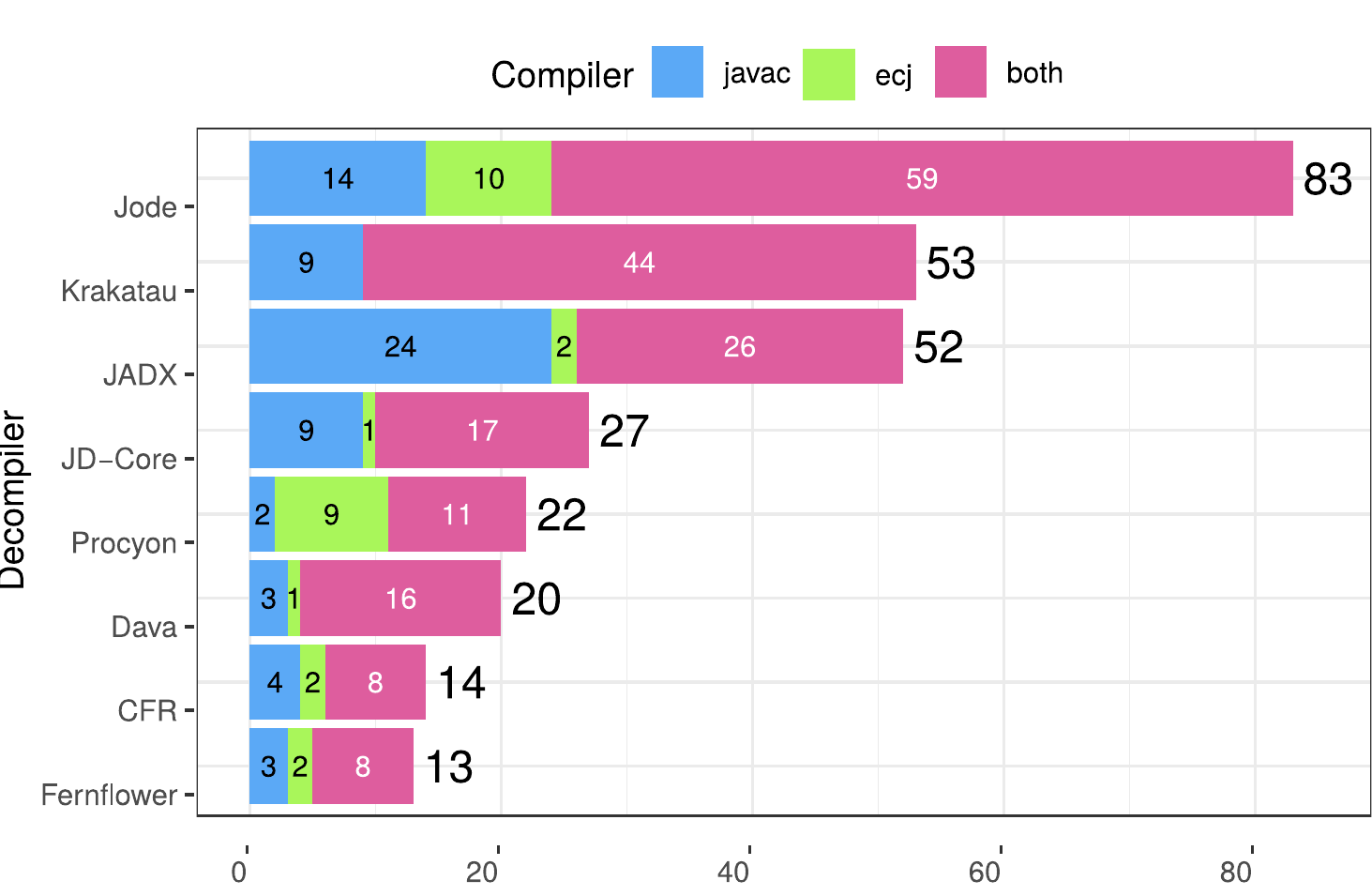}
	\caption{Deceptive decompilation results per decompiler. From left to right, deceptive decompilation occurring after \javac compilation, \ecj and both.}
	\label{fig:r_not_t_compilers}
\end{figure}

\autoref{fig:r_not_t_compilers} shows the distribution of bytecode classes that are deceptively decompiled. Each horizontal bar groups deceptive decompilation per decompiler. The color indicates which compiler was used to produce the class file triggering the error. In blue is the number of classes leading to a decompilation error only when compiled with \javac, in green only when compiled with \ecj, and in pink is the number of classes triggering a decompilation error with both compilers. The sum of these classes is indicated by the total on the right side of each bar.
Note that the bars in \autoref{fig:r_not_t_compilers} represent the number of bug manifestations, which are not necessarily distinct bugs: the same decompiler bug can be triggered by different class files from our benchmark. Also, \autoref{fig:r_not_t_compilers} plots the same classes referred as \texttt{Deceptive} in \autoref{tab:failures}, but in \autoref{tab:failures} classes leading to a deceptive decompilation for both compilers are counted twice.

Overall, \jode is the least reliable decompiler, with 83 decompilation bug instances in our benchmark. While \fernflower produces the least deceptive decompilations on our benchmark ($13$), it is interesting to note that \cfr produces only one more deceptive decompilation ($14$) but that corresponds to fewer bugs per successful decompilation. This makes \cfr the most reliable decompiler on our benchmark.

We manually inspected $10$ of these bug manifestations.
$2$ of them were already reported by other users.
We reported the other $8$ to the authors of decompilers.\footnote{\url{https://github.com/castor-software/decompilercmp/tree/master/funfacts}} The sources of errors include incorrect cast operation, incorrect control-flow restitution, auto unboxing errors, and incorrect reference resolution. Below we detail two of these bugs.

\subsubsection{Case study: incorrect reference resolution}
We analyze the class \texttt{org.bukkit.Bukkit} from the Bukkit project. An excerpt of the original Java source code is given in \autoref{lst:bukkit-original-decompiled}. The method \texttt{setServer} implements a setter of the static field \texttt{Bukkit.server}. This is an implementation of the common Singleton design pattern.
In the context of method \texttt{setServer}, \texttt{server} refers to the parameter of the method, while \texttt{Bukkit.server} refers to the static field of the class \texttt{Bukkit}.

When this source file is compiled with \textit{javac}, it produces a file \texttt{org/bukkit/Bukkit.class} containing the bytecode translation of the original source.
\autoref{lst:bukkit-bytecode} shows an excerpt of this bytecode corresponding to the \texttt{setServer} method (including lines are filled in red, while excluding lines are filled in green)

\begin{lstlisting}[style=Java, float, floatplacement=H, caption={Exerpt of differences in \texttt{org.bukkit.Bukkit} original (in red marked with a -) and decompiled with \jadx sources (in green marked with a +).}, label={lst:bukkit-original-decompiled},belowskip=-6\baselineskip] 
public final class Bukkit {
    private static Server server;
	[...]
    public static void setServer(Server server) {
`-`µaaaaaµ`if (Bukkit.server != null) {`
£+£¤aaaaa¤£if (server != null) {£
            throw new UnsupportedOperationException(
            "Cannot redefine singleton Server");
        }
`-`µaaaaaµ`Bukkit.server = server;`
£+£¤aaaaa¤£server = server;£
        [...]
    }
\end{lstlisting}

When using the \jadx decompiler on \texttt{org/bukkit/-\\Bukkit.class} it produces decompiled sources, of which an excerpt is shown in \autoref{lst:bukkit-original-decompiled}.
In this example, the decompiled code is not semantically equivalent to the original version. Indeed, inside the \texttt{setServer} method the references to the static field \texttt{Bukkit.server} have been simplified into \texttt{server} which is incorrect in this scope as the parameter \texttt{server} overrides the local scope. In the bytecode of the recompiled version (\autoref{lst:bukkit-bytecode}, including lines are filled in green), we can observe  that instructions accessing and writing the static field (\texttt{GETSTATIC}, \texttt{PUTSTATIC}) have been replaced by instructions accessing and writing the local variable instead (\texttt{ALOAD}, \texttt{ASTORE}).

\begin{lstlisting}[style=Java, float, floatplacement=H, caption={Exerpt of bytecode from class \texttt{org/\allowbreak{}bukkit/\allowbreak{}Bukkit.class} compiled with \javac: Lines in red, marked with a -, are in the original bytecode, while lines in green, marked with a +, are from the recompiled sources (decompiled with \jadx).}, label={lst:bukkit-bytecode},belowskip=-6\baselineskip] 
public static setServer(Lorg/bukkit/Server;)V
`-`µaaaµ`GETSTATIC org/bukkit/Bukkit.server :`
`-`µaaaµ`Lorg/bukkit/Server;`
£+£¤aaa¤£ALOAD 0£
     IFNULL L0
     NEW java/lang/UnsupportedOperationException
     DUP
     ATHROW
    L0
     ALOAD 0
`-`µaaaµ`PUTSTATIC org/bukkit/Bukkit.server :`
`-`µaaaµ`Lorg/bukkit/Server;`
£+£¤aaa¤£ASTORE 0£
     ALOAD 0
     INVOKEINTERFACE org/bukkit/Server.getLogger ()Ljava/util/logging/Logger; (itf)
     NEW java/lang/StringBuilder
\end{lstlisting}

When the test suite of \textit{Bukkit} runs on the recompiled bytecode, the $11$ test cases covering this code fail, as the first access to \texttt{setServer} will throw an exception instead of normally initializing the static field \texttt{Bukkit.server}. This is clearly a bug in \jadx.

\subsubsection{Case study: Down cast error}

\begin{lstlisting}[style=Java, float, floatplacement=H, caption={Excerpt of differences in \texttt{FastDatePrinter} original (in red marked with a -) and decompiled with \procyon sources (in green marked with a +).}, label={lst:FastDatePrinter-procyon},belowskip=-6\baselineskip]
    protected StringBuffer applyRules(final Calendar calendar, final StringBuffer buf) {
`-`µaaaaaµ`return (StringBuffer) applyRules(calendar,`
`-`µaaaaaaaaaaaaaaaaaaaaµ`(Appendable) buf);`
£+£¤aaaaa¤£return this.applyRules(calendar, buf);£
    }
    
    private <B extends Appendable> B applyRules(final Calendar calendar, final B buf) {...}
\end{lstlisting}

\autoref{lst:FastDatePrinter-procyon} illustrates the differences between the original sources of \texttt{org/\allowbreak{}apache/\allowbreak{}commons/\allowbreak{}lang3/\allowbreak{}time/\allowbreak{}FastDatePrinter} and the decompiled sources produced by \procyon. The line in red is part of the original, while the line in green is from the decompiled version.
In this example, method \texttt{applyRules} is overloaded, i.e. it has two implementations: one for a \texttt{StringBuffer} parameter and one for a generic \texttt{Appendable} parameter (\texttt{Appendable} is an interface that \texttt{StringBuffer} implements). The implementation for \texttt{StringBuffer}   down casts \texttt{buf} into \texttt{Appendable}, calls the method handling \texttt{Appendable} and casts the result back to \texttt{StringBuffer}. 
In a non-ambiguous context, it is perfectly valid to call a method which takes  \texttt{Appendable} arguments on an instance of a class that implements that interface. But in this context, without the down cast to \texttt{Appendable}, the Java compiler will resolve the method call \texttt{applyRules} to the most concrete method. In this case, this will lead \texttt{applyRules} for \texttt{StringBuffer} to call itself instead of the other method. When executed, this will lead to an infinite recursion ending in a StackOverflowError. Therefore, in this example, \procyon changes the behaviour of the decompiled program   and introduces a bug in it.

\begin{mdframed}[style=mpdframe]
\textbf{Answer to RQ3:} 
Our empirical results indicate that no decompiler is free of deceptive decompilation bugs. 
The developers of decompilers may benefit from the equivalent modulo input concept to find bugs in the wild and extend their test base.
Two bugs found during our study have already been fixed by the decompiler authors, and three others have been acknowledged.

\looseness=-1
\end{mdframed}

\subsection{\textbf{RQ4: (ASTs difference)} \RQfour}

The quality of decompilation depends not only on its syntactic correctness and semantic equivalence but also on how well a human can understand the behaviour of the decompiled program. The code produced by a decompiler may be syntactically and semantically correct but yet hard to read for a human.
In this research question, we evaluate how far the decompiled sources are from the original code.
We measure the syntactic distortion between the original and the decompiled sources as captured by AST differences (Definition \ref{def:syntactic-distortion}). 

\begin{figure}
	\centering
	\includegraphics[origin=c,width=0.49\textwidth]{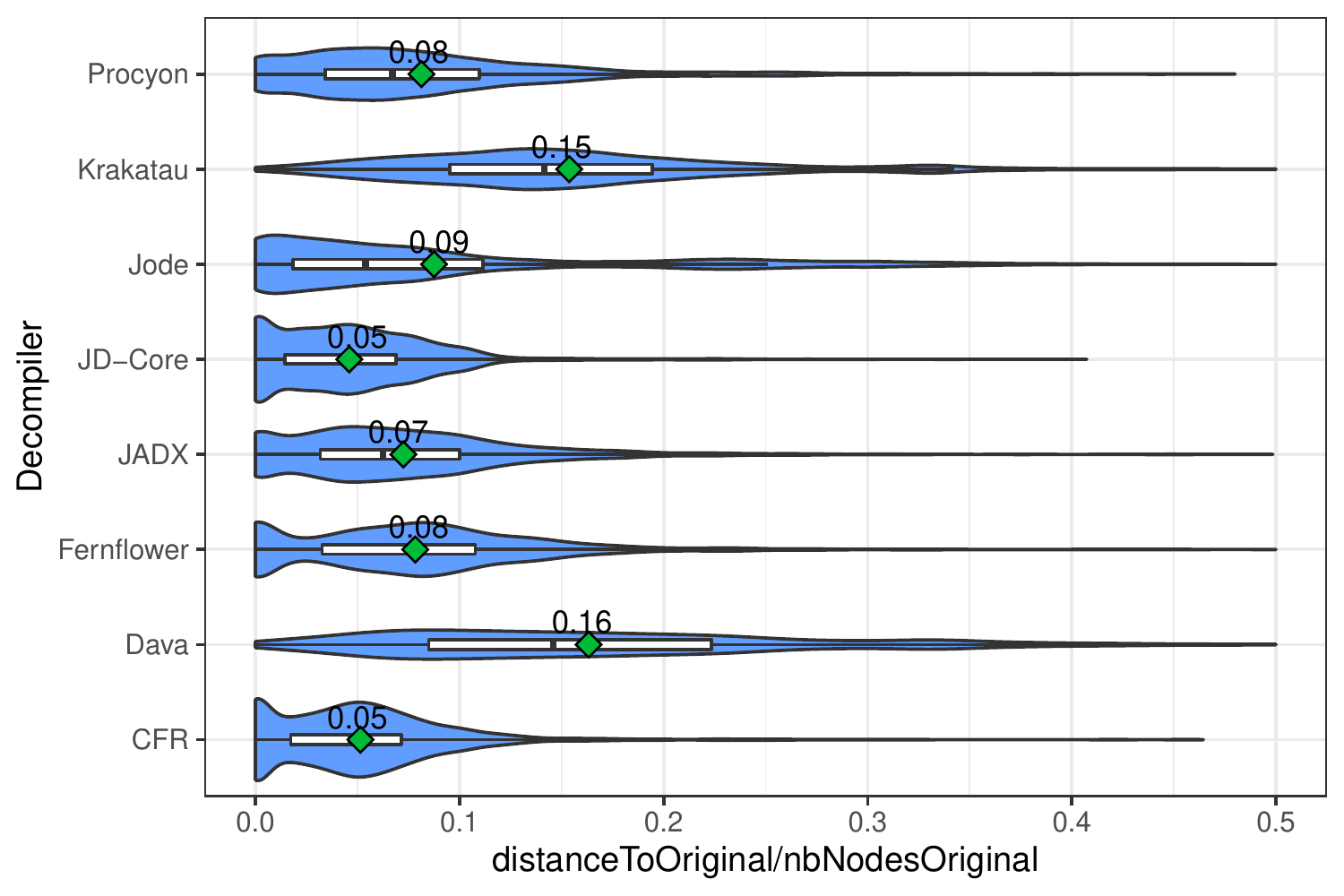}
	\caption{Distribution of  differences between the original and the decompiled source code ASTs. Green diamonds indicate average.}
	\label{fig:distances}
	\vspace{-0.5cm}
\end{figure}

\autoref{fig:distances} shows  the distribution of syntactic distortion present in syntactically correct decompiled code, with one violin plot per decompiler. The green diamond marks the average syntactic distortion. For example, the syntactic distortion values of the \jode decompiler have a median of $0.05$, average of $0.09$, 1st-Q and 3rd-Q of $0.01$ and $0.11$, respectively. In this figure, lower is better: a lower syntactic distortion means that the decompiled sources are more similar to their original counterparts.

\cfr and \jd introduce the least syntactic distortion, with high proportion of cases with no syntactic distortion at all (as we exclude renaming). Their median and average syntactic distortion are close to $0.05$, which corresponds to $5$ edits every $100$ nodes in the AST of the source program. On the other extreme, \dava and \krakatau introduce the most syntactic distortion with average of $16$ (resp. $15$) edits per $100$ nodes. They also have almost no cases for which they produce sources with no syntactic distortion. It is interesting to note that \dava makes no assumptions on the source language nor the compiler used to produce the bytecode it decompiles. \cite{Miecznikowski2002} This partly explains the choice of its author to not reverse some optimizations made by Java compilers (See example introduced in \autoref{sec:background}.).

\autoref{lst:foo-original-fernflower} shows the differences on the resulting source code after decompiling the \texttt{Foo} class from \texttt{DcTest} with \fernflower. As we can observe, both Java programs represent a semantically equivalent program. Yet, their ASTs contain substantial differences. For this example, the edit distance is $3/104$ as it contains three tree edits: \texttt{MOVE} the \texttt{return} node, and \texttt{DELETE} the \texttt{break} node and the \texttt{continue} node (the original source's AST contained $104$ nodes). 

Note that some decompilers perform some transformations on the sources they produce on purpose to increase readability. Therefore, it is perfectly normal to observe some minimal syntactic distortion, even for decompilers producing readable sources. But as our benchmark is composed of non obfuscated sources, it is expected that a readable output will not fall too far from the original.\\

\textbf{}\begin{lstlisting}[style=Java, float, floatplacement=H, caption={Excerpt of differences in \texttt{Foo} original (in red marked with a -) and decompiled with \fernflower  (in green marked with a +) sources.}, label={lst:foo-original-fernflower},belowskip=-6\baselineskip]
public class Foo {
  public int foo(int i, int j) {
    while (true) {
      try {   
        while (i < j) i = j++ / i;
£+£¤aaaaa¤£return j;£
      } catch (RuntimeException re) {
        i = 10;
`-`µaaaaaµ`continue;`
      }
`-`µaaaµ`break;`                   
    }
`-`µaaµ`return j;`
  }
}
\end{lstlisting}

	\vspace{-0.5cm}
\begin{mdframed}[style=mpdframe]
\textbf{Answer to RQ4:} 
All decompilers present various degrees of  syntactic distortion between the original source code and the decompiled bytecode. This reveals that all decompilers adopt different strategies to craft source code from bytecode. 
We propose a novel metric to quantify the distance between the original source code and its decompiled counterpart.
Also, decompiler users can use this analysis when deciding which decompiler to employ.
\looseness=-1
\end{mdframed}

\subsection{\textbf{RQ5: (Decompiler Diversity) \RQfive}}
\label{sec:rq5}

In the previous research questions, we observe that different decompilers produce source code that varies in terms of syntactic correctness, semantic equivalence and syntactic distortion. Now, we investigate the overlap in successes and failures of the different decompilers considered for this study.

\begin{figure}
	\centering
	\includegraphics[origin=c,width=0.42\textwidth]{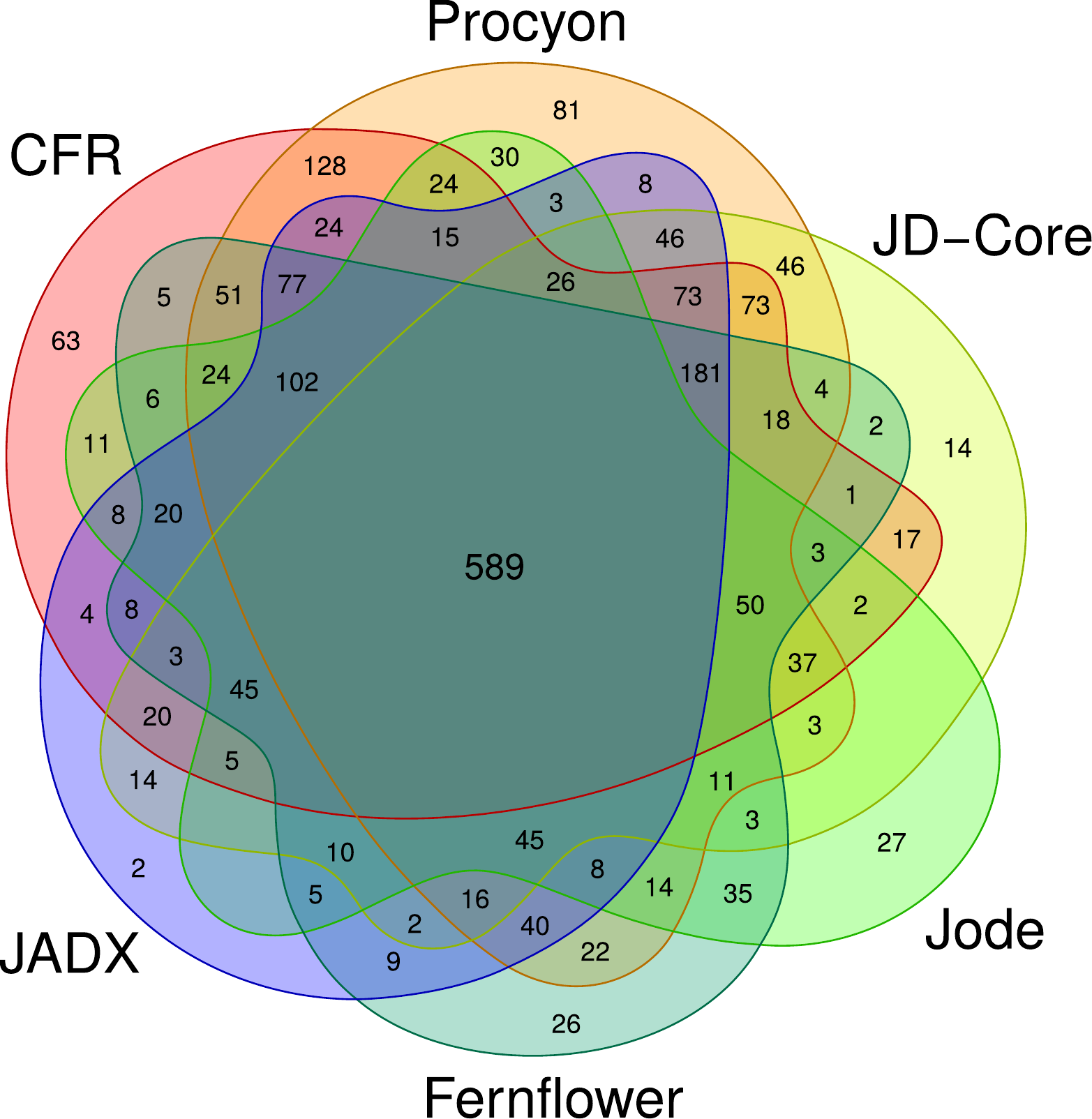}
	\caption{Venn diagram of syntactically and semantically equivalent modulo inputs decompilation results.}
	\label{fig:venn_diagram_recompilable}
	\vspace{-0.5cm}
\end{figure}

\autoref{fig:venn_diagram_recompilable} shows a Venn Diagram of semantically equivalent classes modulo inputs for decompiled/recompiled classes. We exclude \dava and \krakatau because they do not handle correctly any class file unique to them.
We see that 6/8 decompilers have cases for which they are the only decompiler able to handle it properly. These cases represent $276/2397$ classes. Only $589/2397$ classes are handled correctly by all of these 6 decompilers. Furthermore, $157/2397$ classes are not correctly handled by any of the considered decompilers.

\textbf{}\begin{lstlisting}[style=Java, float, floatplacement=H, caption={Excerpt of differences in \texttt{AbstractHashedMap} original (in red marked with a -) and decompiled with \procyon  (in green marked with a +).}, label={lst:keysetiterator},belowskip=-6\baselineskip]
protected static class KeySetIterator<K> extends HashIterator<K, Object> implements Iterator<K> {
`-`µaaµ`@SuppressWarnings("unchecked")`
        protected KeySetIterator(final AbstractHashedMap<K, ?> parent) {
`-`µaaaaµ`super((AbstractHashedMap<K, Object>) parent);`
£+£¤aaaa¤£super(parent);£
        }
    [...]
}
\end{lstlisting}
\autoref{lst:keysetiterator} is an excerpt of \texttt{AbstractHashedMap}, which is incorrectly decompiled by all decompilers. While the complete set of syntactic errors for the decompiled sources is different for each decompiler, it always includes one call of the constructor of the super class of \texttt{KeySetIterator<K>}. Either a constructor with the correct signature is not resolved or the cast in front of \texttt{parent} is missing. The fundamental problem behind this decompilation lies in the fact that the JVM does not directly support generics\cite{javabrocken}. While bytecode do keep meta information about types in signatures, the actual type manipulated in this example for \texttt{?} is an \texttt{Object}. Therefore, contrarily to the original sources, no \texttt{CHECKCAST} instruction is required in the bytecode. This does not make the task of decompilation impossible to perform in theory, as both the type of \texttt{parent} and the type required by the \texttt{super} constructor can be resolved, but, it does make it more challenging to decompilers in practice.

A manual analysis of these classes shows  common issues among the studied decompilers. (i) Generics is a feature that causes many decompilers to fail in particular when combined with ternary operators, wildcards or type bounds. Another example of such a case is detailed in \autoref{sec:meta}. (ii) As mentioned in \autoref{sec:rq1-results} and  \autoref{sec:rq3-results}, compilers producing the bytecode do play a role. In particular, synthetic elements created by a compiler,  which the decompiler does not expect. (iii) Overall, the diversity of independent corner cases cannot be completely captured under one concise explanation. Even for \procyon, the best performing decompiler in our study,  among the $528$ classes for which it does not produce semantically equivalent modulo inputs sources, only $157$ are also not decompilable by any other decompiler.

\begin{table}
    \centering
    \scriptsize
    \caption{Summary results of the studied decompilers}
    \begin{tabular}{lcccccc}
      \hline
     \textsc{\textbf{Decompiler}} & \textsc{\textbf{\#Recompilable}} & \textsc{\textbf{\#PassTest}} & \textsc{\textbf{\#Deceptive}} \\
      \hline
      \cfr & $3097$ ($0.79$) & $1713$ ($0.71$) &  $22$ \\
      \dava & $1747$ ($0.44$) & $762$ ($0.32$) &  $36$ \\
      \fernflower & $2663$ ($0.68$) & $1435$ ($0.60$) &  $21$ \\
      \jadx & $2736$ ($0.70$) & $1408$ ($0.59$) &  $78$ \\
      \jd & $2726$ ($0.69$) & $1375$ ($0.57$) &  $44$ \\
      \jode & $2569$ ($0.65$) & $1161$ ($0.48$) & $142$ \\
      \krakatau & $1746$ ($0.44$) & $724$ ($0.30$) &  $97$ \\
      \procyon & $3281$ ($0.84$) & $1869$ ($0.78$) &  $33$ \\
      \hline
      Union & $3734$ ($0.95$) & $2240$ ($0.93$) &  $342$ \\
      \hline
      Total & $3928$ ($1.00$) & $2397$ ($1.00$) &  - \\
      \hline
    \end{tabular}
    \label{tab:overall}
	\vspace{-0.5cm}
\end{table}

\autoref{tab:overall} summarizes the quantitative results obtained from the previous research questions. Each line corresponds to a decompiler. Column \texttt{\#Recompilable} shows the number of cases (and ratio) for which the decompiler produced a recompilable output among all classes of our dataset ($3928$ in total: $2041$ for \javac and $1887$ for \ecj). Column \texttt{\#PassTest} shows the number of cases where the decompiled code passes those tests among the $2397$ classes covered by tests and regrouping both compiler. Column \texttt{\#Deceptive} indicates the number of cases that were recompilable but did not pass the test suite (i.e. a decompilation bug). 
The line `Union` shows the number of classes for which at least $1$ decompiler succeeds to produce \texttt{Recompilable} sources and respectively sources that \texttt{pass tests}. The column \texttt{\#Deceptive} indicates the number of classes for which at least $1$ decompiler produced a deceptive decompilation.
This means that for $2240$ classes out of the $2397$ ($93\%$), there is at least $1$ decompiler that produces semantically equivalent sources modulo inputs. This number must be taken with a grain of salt, as it does not mean that someone who looks for a successful decompilation of one of these classes could find one trivially. Overall, $342$ out of $2397$ classes have at least $1$ decompiler that produce a deceptive decompilation. Assuming that one can merge the successful decompilation results together, we would obtain a better decompiler overall, this is what we explore in \autoref{sec:meta}.

\newpage
\begin{mdframed}[style=mpdframe]
\textbf{Answer to RQ5:} 
The classes for which each decompiler produce semantically equivalent source code modulo input do not overlap completely. For $6$ out of $8$ decompilers, there exists at least $1$ class for which the decompiler is the only one to produce \semi sources.
In theory, a union of the best features of each decompiler would cover $2240$ out of the $2397$ ($93\%$) classes of the dataset. This suggests to combine multiple decompilers  to improve decompilation results. 
\end{mdframed}

\section{\textbf{Meta Decompilation}}
\label{sec:meta}

In this section, we present an original concept for decompilation.

\subsection{Overview}

In 1995, Selberg et al.~\cite{metacrawler} noticed that different web search engines produced different results for the same input query. They exploited this finding in a tool called \textsc{MetaCrawler}, which delegates a user query to various search engines and merges the results. 
This idea of combining diverse tools that have the same goal has been explored since then.
For example, Blair and Somayaji \cite{foster2010object} explore how a genetic algorithm can recombine related programs at the object
file level to produce correct variants of C programs.
Persaud et al. \cite{persaud2016frankenssl} combines cryptographic libraries together for software security. Chen et al.~\cite{Chen2018} rely on various fuzzers to build an ensemble based fuzzer that gets better performance and generalization ability than that of any constituent fuzzer alone.

In this paper, we apply a similar approach to improve Java decompilation. Each decompiler has its strengths and weaknesses, and the subset of JVM bytecode sequences they correctly handle is diverse (cf \autoref{sec:rq5}). Therefore, our idea is to combine decompilers in a meta-decompiler. 

In this paper, we propose a tool called \metadc, that implements such a `meta-decompilation' approach. \metadc merges partially incorrect decompilation results from diverse decompilers in order to produce a correct one.

\subsection{Example}

The class \texttt{org.\allowbreak{}bukkit.\allowbreak{}configuration.\allowbreak{}file.\allowbreak{}YamlConfiguration} of the project Bukkit is an example of a class file that is incorrectly handled by both \jadx and \dava. While both decompilers produce syntactically incorrect Java code for this class, the error that prevents successful recompilation is not located at the same place in both decompiled classes.

\textbf{}\begin{lstlisting}[style=Java, float, floatplacement=H, caption={Excerpt of \texttt{org.bukkit.configuration. file.YamlConfiguration} decompiled with \dava.}, label={lst:yaml-dava}, escapechar=\¤,belowskip=-6\baselineskip]
public class YamlConfiguration extends FileConfiguration {
    protected static final String COMMENT_PREFIX = "# ";
    protected static final String BLANK_CONFIG = ¤\highlightred{\textcolor{pgreen}{"\{\}}}¤
¤\highlightred{\textcolor{pgreen}{"};}¤
    private final DumperOptions yamlOptions;
    private final Representer yamlRepresenter;
    private final Yaml yaml;

    public YamlConfiguration()
    {
        DumperOptions r7;
        YamlRepresenter r8;
        YamlConstructor r9;
        Yaml r10;
        BaseConstructor r11;
        r7 = new DumperOptions();
        yamlOptions = r7;
        r8 = new YamlRepresenter();
        yamlRepresenter = r8;
        r9 = new YamlConstructor();
        r11 = (BaseConstructor) r9;
        r10 = new Yaml(r11, yamlRepresenter, yamlOptions);
        yaml = r10;
    }
  [...]
}
\end{lstlisting}

\textbf{}\begin{lstlisting}[style=Java, float, floatplacement=H, caption={Excerpt of \texttt{org.bukkit.configuration. file.YamlConfiguration} decompiled with \jadx.}, label={lst:yaml-jadx},belowskip=-6\baselineskip]
public class YamlConfiguration extends FileConfiguration {
    protected static final String BLANK_CONFIG = "{}\n";
    protected static final String COMMENT_PREFIX = "# ";
    private final Yaml yaml = 
        new Yaml(new YamlConstructor(), 
        `this.yamlRepresenter, this.yamlOptions);`
    private final DumperOptions yamlOptions = new DumperOptions();
    private final Representer yamlRepresenter = new YamlRepresenter();
  [...]
}
\end{lstlisting}

\autoref{lst:yaml-dava} shows an excerpt of the decompiled sources produced by \dava for \texttt{YamlConfiguration}. The static field \texttt{BLANK\_CONFIG} is initialized with an incorrect string literal that contains a non escaped line return. When attempting to recompile these sources, \javac produces an \texttt{unclosed string literal} error for both line 3 and 4.

\autoref{lst:yaml-jadx} shows an excerpt of the decompiled sources produced by \jadx for the same class. The static field \texttt{BLANK\_CONFIG} is correctly initialized with \texttt{"\{\}\textbackslash n"}, but the initialization of \texttt{yaml}, \texttt{yamlOptions} and \texttt{yamlRepresenter} are conducted out of order, which lead to a compilation error as \texttt{yamlOptions} and \texttt{yamlRepresenter} are still \texttt{null} when \texttt{yaml} is initialized.
Intuitively, one can see that \dava's solution could be fixed by replacing lines 3 and 4 with line 2 from \jadx's solution. This is an example of successful meta-decompilation, merging the output of two decompilers.

\begin{figure}
	\centering
	\includegraphics[origin=c,width=0.48\textwidth]{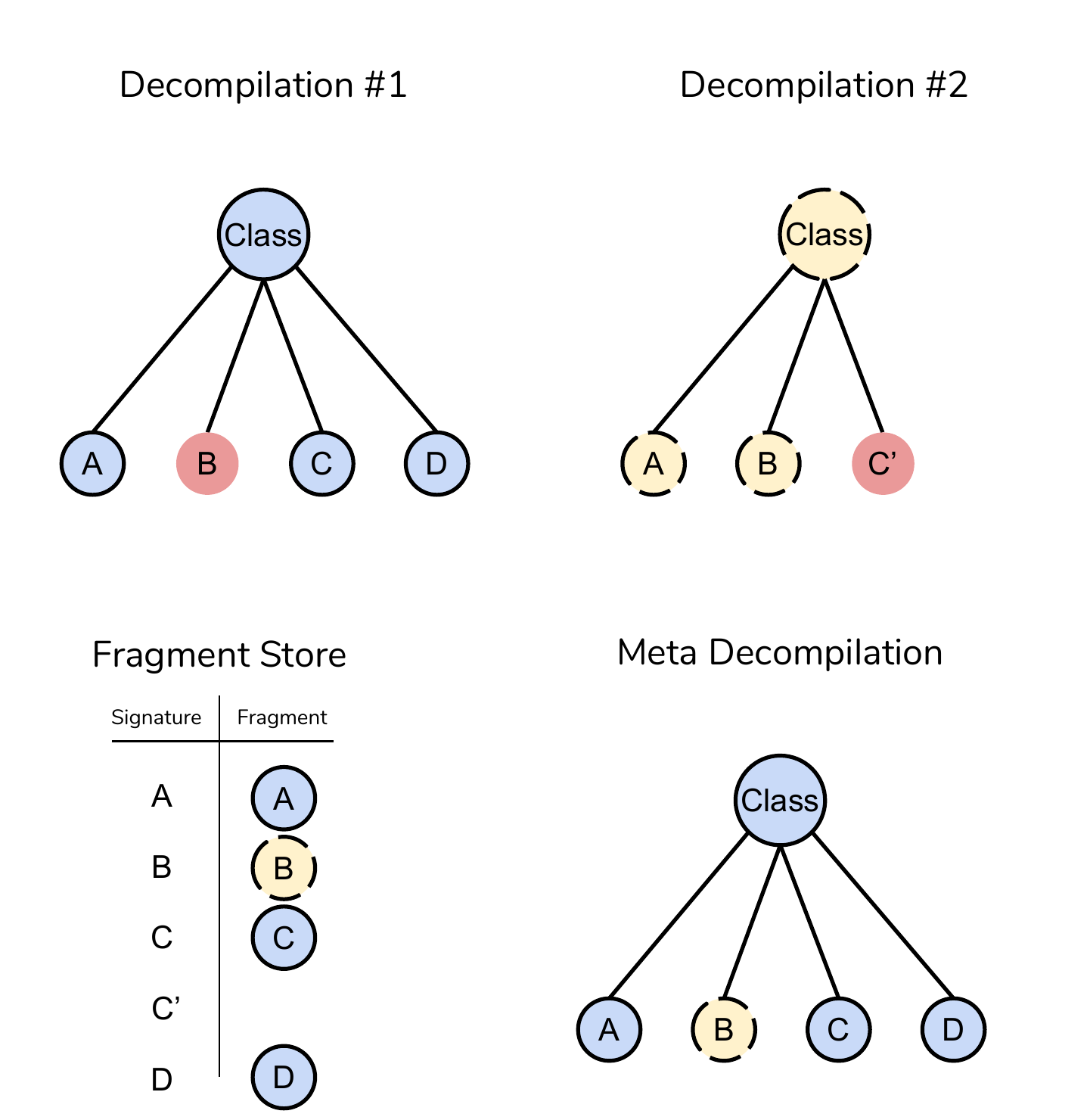}
    \caption{Meta decompilation: Merger of different partial decompilation.
    Node in blue \modified{with plain border} originates from Decompiler \#1, nodes in yellow with dashed border originate from Decompiler \#2. Borderless nodes in red contains compilation errors.}
	\label{fig:meta-step}
\end{figure}
\autoref{fig:meta-step} illustrates how two erroneous decompilations can be merged into one that is correct, when the error is not located at the same place.
This figure represents different versions of the abstract syntax tree (AST) of a Java class. The root node corresponds to the class itself, while its children represent type members of the class. A type member is either a method, a field, a nested type (class, or enum), or a static initialization block.
\texttt{Decompilation \#1} represents the AST of the sources produced by one decompiler, it includes $4$ type members (A,B,C and D), and one compilation error located in B. \texttt{Decompilation \#2} represents the sources produced by a different decompiler for the same class. It  contains only $3$  type members (A,B an C') and one compilation error located in C'. The \store is a dictionary containing an error free AST fragment for each type member when such a fragment exists. \texttt{Meta Decompilation} shows an example of error free AST that can be built based on \texttt{Decompilation \#1} and the store that combines AST fragments from both decompilations. Note that different decompilers may produce sources that do not exactly contain the same type members. This is illustrated here by \texttt{Decompilation \#2} not having a type member D and having a different signature for C.

\subsection{Algorithm}

\begin{algorithm}[htb]
  \SetAlgoLined
  \SetNoFillComment
  \KwData{$bytecode$ A bytecode file, \\$Decompilers$ A set of decompilers} 
  \KwResult{The decompiled java sources corresponding}
  
  $Solutions \gets \{\}$
  
  $FragmentStore \gets \{\}$

  \ForEach{$dc \in Decompilers$}{
    $solution \gets AST(decompile(dc, bytecode))$\label{line:dc}
    
    $Fragments \gets fragmentsOf(solution)$
    
    \ForEach{$f \in Fragments$}{
      \If{$\neg problem(f) \land signature(f) \notin Store$}{
        $FragmentStore \gets FragmentStore \cup \{signature(f) \to f\}$\label{line:add-to-store}
      }
    }
    
    $Solutions \gets Solutions \cup \{solution$\}
    
    \ForEach{$s \in Solutions$}{\label{line:check-sol}
      \If{$completable(s,FragmentStore)$}{
        \eIf{$recompile(complete(s,FragmentStore))$}{
          \Return $print(s)$\;
        }{
          $remove(s,FragmentStore)$
        }
      }
    }
  }

\caption{Meta decompilation procedure.}
\label{alg:meta-algo}
\end{algorithm}

\autoref{alg:meta-algo} describes the process of meta decompilation as implemented by \metadc. \metadc takes as input a bytecode file, and an ordered list of bytecode decompilers. The process starts with an empty set of solutions and an empty \store of correct fragments. 
This \store is a dictionary that associates a type member signature to a fragment of AST free of compilation error corresponding to the type member in question.

For each decompiler, the meta-decompilation goes through the following steps.

The bytecode file is passed to the decompiler $d$. An AST is built from the decompiled sources (\autoref{line:dc}). While building the AST, the compilation errors and their location are gathered (if any) and the type members containing errors are annotated as such. A class abstract syntax tree includes a node for the class itself as the root, as well as children representing class information (super class, super interfaces, formal type parameters) and type members. Type members include fields, methods, constructors, inner classes, enum values, and static blocks. These type members' source locations are recorded and compared with the compiler error locations. If an error is located between a type member start and end location, the type member is annotated as errored. For example, the element corresponding to the field \texttt{BLANK\_CONFIG} is  annotated as errored in \dava's solution for \texttt{YamlConfiguration}.
This annotated AST, that we call solution, is added to the set of remaining solutions.

Additionally, for all type members in the current solution, if the \store does not already contain an error free fragment with the same signature, the type member is added to the \store (\autoref{line:add-to-store}).
The signature of a type member is a character string that identifies it uniquely.
For example, the signature of the field \texttt{BLANK\_CONFIG} is \texttt{org.\allowbreak{}bukkit.\allowbreak{}configuration.\allowbreak{}file.\allowbreak{}YamlConfiguration\#\allowbreak{}BLANK\_CONFIG} and the signature of \texttt{YamlConfiguration}'s constructor is \texttt{org.\allowbreak{}bukkit.\allowbreak{}configuration.\allowbreak{}file.\allowbreak{}YamlConfiguration()}.

Each solution in the set of solutions is checked for completion with the current store (\autoref{line:check-sol}). A solution is ``completable'' with the members in a given \store, if all the solution's type members annotated with an error are present in the \store. Indeed, these type members' AST can be replaced with an error free variant present in the \store.
If a solution is completable with the current \store, all its type members annotated as errored are replaced with a fragment from the \store.
The solution is then passed to the compiler to check if it compiles. If it does, it is printed, and the meta decompilation stops. If not, the solution is removed from the set of solutions. As the first solution that satisfies the oracle (syntactic correctness) stops the process, and this oracle is imperfect, the order of the decompilers matters. More details are given in the following section.

By attempting to repair each solution and its given set of type members with a minimum of transplanted fragments from those available in the \store, \metadc does not favour any type member set.
This allows \metadc to deal with cases where the different solutions do not contain the same type members. 
This occurs with implicit constructor declarations such as the one present in \autoref{lst:yaml-jadx} with \texttt{YamlConfiguration}. It also makes it possible to handle cases where element signatures might differ depending on how type erasure is dealt with by each decompiler.
And finally, it handles cases where elements might not be in the same order (and the order of type members is meaningful as seen in \autoref{lst:yaml-jadx}).

\subsection{Experimental results about meta-decompilation}

The following section evaluates the effectiveness of \metadc. It is organized as follows.
First, we gather the $157$ classes of our dataset for which no decompilers produced semantically equivalent modulo input sources and assess the results produced by \metadc.
Second, we run \metadc on the complete dataset of classes in this study. We then evaluate the results with regards to semantic equivalence modulo inputs.
Finally, we study the origin of fragments produced by \metadc and discuss the consequences on the number of deceptive decompilations.

\begin{table}
\centering
\caption{\metadc results on classes with no correct decompilation from state of the art decompilers.}
\footnotesize
\begin{tabular}{l|ccc}
            & \textsc{\metadc} & Union\\
            \hline
    		\#PassTest & 59 (37.6\%) & 0 (0\%)\\
    		\#Deceptive & 11 (7.0\%) & 23 (14.6\%)\\
    		\#!Recompile & 87 (55.4\%) & 134 (85.4\%)\\
            \hline
    		Total & 157 (100\%) & 157 (100\%)\\
            \hline
		\end{tabular}
	\label{tab:meta-problem}
\end{table}
\autoref{tab:meta-problem} shows the results of meta decompilation on the $157$ classes of our dataset that led to decompilation errors for all decompilers in the study and were covered by at least one test\footnote{3 of the 137 classes that led to syntactically incorrect outputs for all decompilers are not covered by any tests.}. \metadc produces semantically equivalent results for $59$ out of $157$ ($37.6\%$) classes. It produces deceptive decompilation for $11$ ($7.0\%$) classes and fails to produce recompilable results for $87$ out of $157$ ($55.4\%$) classes.
The success case where \metadc produces correct output is when:
1) at least one compiler is able to read the correct signature for all type members of a class and, 
2) an error free decompilation exists for all of these type members.
However, when no decompiler is able to decompile a specific type member or that no decompiler reads correctly the signature of all type members, no meta decompilation can be successful.
These results demonstrate that successful decompilation (in terms of both syntactic correctness and \semi) can be found by \metadc  for classes where no other decompilers can.

\begin{table}
    \centering
    \scriptsize
    \caption{Comparison of \metadc results with state of the art.}
    \begin{tabular}{lccccccc}
     \textsc{\textbf{Decompiler}} &
     \rotatebox{60}{\textsc{\textbf{\#Recompilable}}} &
     \rotatebox{60}{\textsc{\textbf{\#PassTest}}} &
     \rotatebox{60}{\textsc{\textbf{\#Deceptive}}} &
     \rotatebox{60}{\textsc{\textbf{ASTDiff}}} \\ 
      \hline
      \cfr & $3097$ ($79\%$) & $1713$ ($71\%$) &  $22$ ($1.27\%$) & $0.05$ \\ 
      \procyon & $3281$ ($84\%$) & $1869$ ($78\%$) &  $33$ ($1.74\%$) & $0.08$ \\ 
      \metadc & $3479$ ($89\%$) & $2087$ ($87\%$) &  $30$ ($1.42\%$) & $0.06$ \\ 
      \hline 
      Total & $3928$ ($100\%$) & $2397$ ($100\%$) &  - & - \\ 
      \hline 
    \end{tabular}
    \label{tab:meta-overall}
\end{table}
\autoref{tab:meta-overall} shows the results obtained when running \metadc on the whole dataset presented in \autoref{sec:methodology} and compares it with \procyon and \cfr. \procyon is the decompiler that scores the highest in terms of syntactic correctness as well as \semi, while \cfr scores the lowest in deceptive decompilation rate and syntactic distortion. 
The first column indicates the number of classes for which each decompiler produced syntactically correct sources, among the $3928$ from the dataset. The second column shows the number of classes for which each decompiler produced semantically correct modulo inputs sources among the $2397$ classes covered by tests. The third column indicates the number of deceptive decompilations produced by each decompiler. The percentage of deceptive decompilation is computed with \texttt{\#Deceptive} $/$ (\texttt{\#Deceptive} $+$ \texttt{\#PassTests}). The last column shows the median syntactic distortion in number of edits per nodes in the original AST.

\metadc produced syntactically correct sources for $3479$ classes ($89\%)$. It produces semantically equivalent modulo inputs sources for $2087$ ($87\%$) classes, and $30$ deceptive decompilations. 
Compared with \procyon, \metadc produces syntactically correct sources for $198$ more classes, semantically correct modulo inputs sources for $218$ more. It also produces $3$ less deceptive decompilations, and has a lower syntactic distortion.
Compared with \cfr, \metadc produces $8$ more deceptive decompilations but it produces semantically correct modulo inputs sources for $374$ more classes. In percentage of deceptive decompilation among recompilable decompilation, \metadc produces $1.42\%$ of deceptive decompilation which is lower than \procyon's $1.74\%$ but slightly higher than \cfr's $1.27\%$.

Overall, \metadc scores higher than all studied decompilers in terms of semantic correctness as well as semantic equivalence modulo inputs, and ranks second in deceptive decompilation rate by a small margin. The rate of semantically equivalent decompilation modulo inputs is higher because \metadc produces, by design, more syntactically correct decompilations. On the other hand, the rate of deceptive decompilation is slightly higher than \cfr, as \metadc aggregates some of the deceptive decompilations from all used decompilers. This is, to our knowledge, the first implementation of this meta-decompilation approach. It  demonstrates the validity of the approach and adds a new state of the art tool that practitioners can use to decompile Java bytecode.

Note that \metadc also has its implementation flaws and may fail where other decompilers may succeed. In particular, not all AST nodes transplantation produce syntactically correct code.
But it may be used in conjunction of other decompilers. The union of classes for which at least one decompiler (including \metadc) produces semantically equivalent modulo inputs sources, presented in RQ5, now covers $2299$ out of $2397$ classes ($96\%$) of our dataset.

\subsubsection{Remaining deceptive decompilations}

In order to investigate deceptive decompilations produced by \metadc, we need to investigate the origins of the AST fragments used in each decompilation.

\begin{figure}
	\centering
	\includegraphics[origin=c,width=0.48\textwidth]{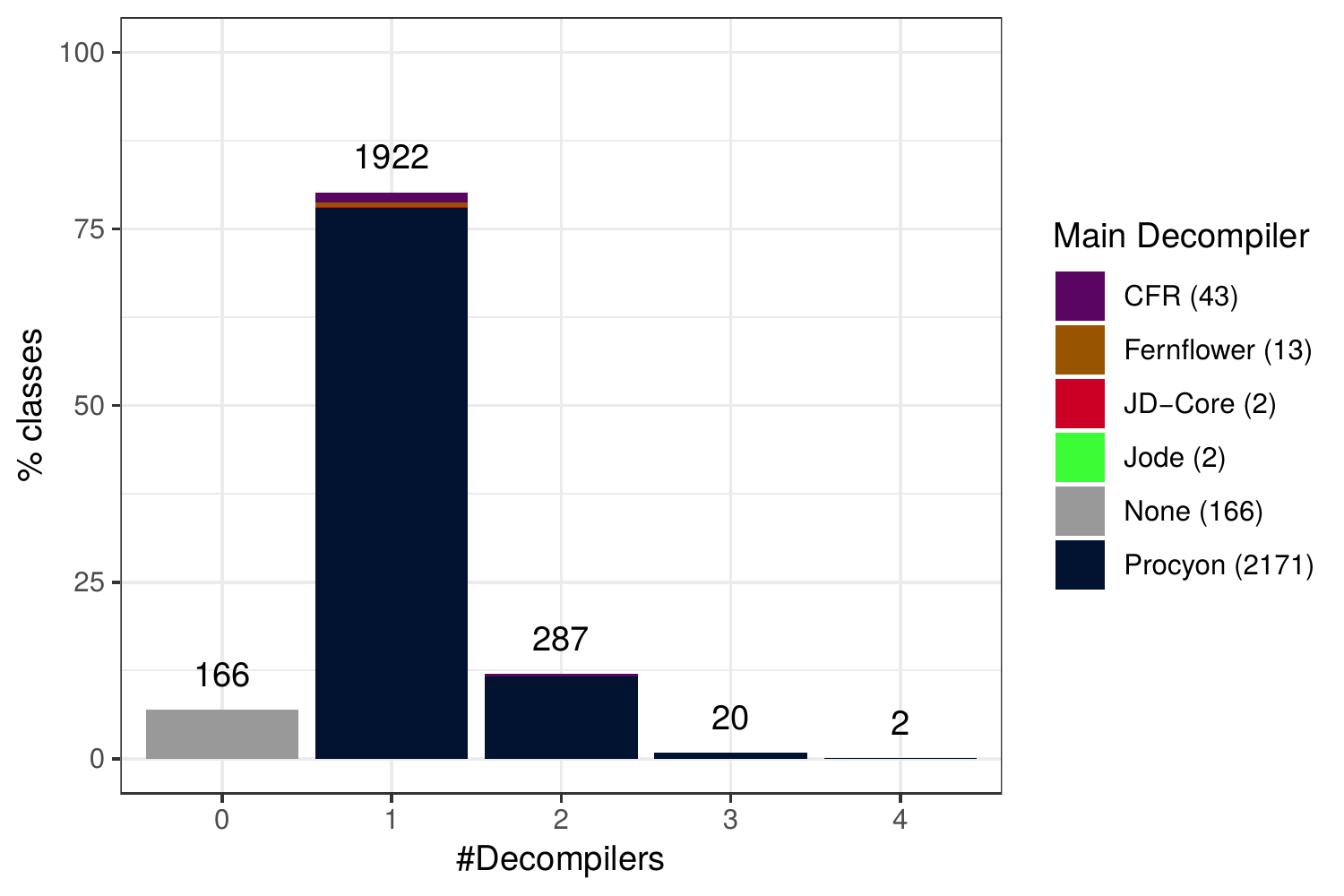}
	\caption{Distribution of the number of decompilers used by \metadc.}
	\label{fig:meta-main}
	\includegraphics[origin=c,width=0.48\textwidth]{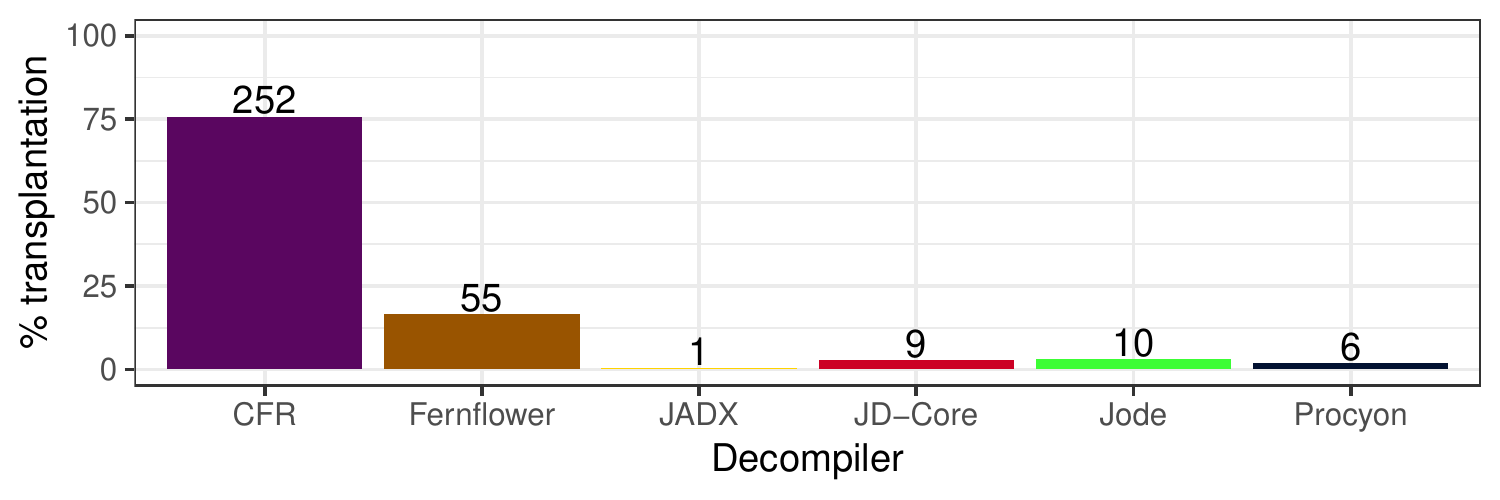}
	\caption{Distribution of the origin of transplanted fragments in \metadc results.}
	\label{fig:meta-transplant}
\end{figure}

\autoref{fig:meta-main} shows the distribution of the number of decompilers used by \metadc for each of the $2397$ classes of our dataset for which we have tests. \metadc finds no solution for $166$ classes. For $1922$ classes, only one decompiler was used, meaning that there is no need for meta-decompilation.
For $287$ classes, \metadc combines the output of $2$ decompilers. It uses $3$ and $4$ decompilers for $20$ classes and $2$ classes respectively. The color indicates which decompiler's base solution was used. In the overwhelming majority, the  \procyon solution is used.

\autoref{fig:meta-transplant} shows the distribution of transplanted fragments' origin for the $309$ classes where several decompilers are used. For $252$ classes, one or more fragments from \cfr's solution were transplanted to build \metadc's solution. $55$ classes have fragments coming from \fernflower, and the rest of the distribution is negligible.
Note that \metadc stops as soon as it finds an admissible solution. Thus, the order of decompilers when building a solution largely impacts this distribution.

\metadc produces a deceptive decompilation either when the first recompilable solution of a given type member is a deceptive one, or the assembly of different fragments introduces an error.

In order to minimize these problems, \metadc uses \procyon as the first decompiler and orders the other decompilers by their deceptive decompilation rate.

Therefore, most of the decompilers' deceptive decompilations are for the same classes as \procyon's one. In a lesser way, deceptive decompilation originating from type members decompiled with \cfr affect \metadc when those type member are decompiled with syntactic errors by \procyon.
Note that, as no software is free of bugs, the implementation of \metadc could also add new sources of error.
In practice, as shown by \autoref{tab:meta-overall}, the number of deceptive decompilations ($30$) corresponds to a better deceptive decompilation rate than all decompilers of this study except \cfr.

\subsubsection{Case studies}

Here we discuss two examples in details: one successful and one failed meta decompilation.

\textbf{}\begin{lstlisting}[style=Java, float, floatplacement=H, caption={Excerpt of \texttt{org.junit.runner.Request} decompiled with \procyon.}, label={lst:request-procyon}, escapechar=\¤,belowskip=-6\baselineskip]
`import org.junit.runners.model.*;`
`import org.junit.internal.runners.*;`
public abstract class Request {
	[...]
    public static Request classes(final Computer computer, final Class<?>... classes) {
        try {
            final AllDefaultPossibilitiesBuilder builder = new AllDefaultPossibilitiesBuilder(true);
            final Runner suite = computer.getSuite(builder, classes);
            return runner(suite);
        }
        catch (¤\highlightred{InitializationError}¤ e) {
            throw new RuntimeException("Bug in saff's brain: Suite constructor, called as above, should always complete");
        }
    }
    
    public static Request runner(final Runner runner) {
        return new Request() {
            @Override
            public Runner getRunner() {
                return runner;
            }
        };
    }
}
\end{lstlisting}
\textbf{}\begin{lstlisting}[style=Java, float, floatplacement=H, caption={Excerpt of \texttt{org.junit.runner.Request} decompiled with \cfr .}, label={lst:request-cfr},belowskip=-6\baselineskip]
import org.junit.runners.model.InitializationError;

public abstract class Request {
	[...]

    public static Request classes(Computer computer, Class<?> ... classes) {
        try {
            AllDefaultPossibilitiesBuilder builder = new AllDefaultPossibilitiesBuilder(true);
            Runner suite = computer.getSuite(builder, classes);
            return Request.runner(suite);
        }
        catch (InitializationError e) {
            throw new RuntimeException("Bug in saff's brain: Suite constructor, called as above, should always complete");
        }
    }

    public static Request runner(Runner runner) {
        return new Request(){

            public Runner getRunner() {
                `return Runner.this;`
            }
        };
    }
}
\end{lstlisting}

\paragraph{Success: Request.} 

\autoref{lst:request-procyon} shows the decompiled sources for \texttt{org.junit.runner.Request} produced by \procyon. In this example, there are ambiguous references because two types share the same simply qualified name: both \texttt{org.\allowbreak{}junit.\allowbreak{}runners.\allowbreak{}model} and \texttt{org.\allowbreak{}junit.\allowbreak{}internal.runners} contain a type named \texttt{InitializationError}, therefore the decompiled sources generated by \procyon lead to a compilation error. 

\autoref{lst:request-cfr} shows the decompiled sources for \texttt{org.\allowbreak{}junit.\allowbreak{}runner.\allowbreak{}Request} produced by \cfr. These sources contain an error in the body of the static method \texttt{runner(Runner)}. Since this method contains an anonymous class, when the original sources are compiled, a synthetic field \texttt{runner} is created, by the compiler, for the anonymous class. This field contains the parameter \texttt{runner} from the enclosing method. When \cfr decompiles the bytecode, it incorrectly replaces the statement that returns the parameter of the enclosing method by a statement that returns a field that does not exist in the sources. This leads to a compilation error when attempting to recompile. Since our report, \cfr's author has fixed this bug.\footnote{\url{https://github.com/leibnitz27/cfr/issues/50}}

While both \procyon and \cfr's solutions contain an error, these errors are not located on the same type member. Hence, \cfr's fragment for the method \texttt{classes(Computer, Class<?>[])} is transplanted on \procyon's solution. Since the pretty printer used by \metadc only lists imports at a type granularity, and  \cfr's fragment contains references that are non-ambiguous, the combined solution is recompilable and semantically equivalent modulo input.

\paragraph{Failure: SwitchClosure.}

There are Java constructs for which all decompilers struggle. In these cases, all decompilers may produce an error on the same type member, and this leads to a failed meta-decompilation. 
The following example illustrates the problem of generic type lower bounds, which challenges all decompilers.

\textbf{}\begin{lstlisting}[style=Java, float, floatplacement=H, caption={Excerpt of \texttt{org.apache.commons. collections4.functors.SwitchClosure}, original (in red marked with a -) and decompiled (in green marked with a +).}, label={lst:switch-closure},belowskip=-6\baselineskip]
    private final Closure<? super E> iDefault;

    private SwitchClosure(final boolean clone, 
        final Predicate<? super E>[] predicates,
        final Closure<? super E>[] closures, 
        final Closure<? super E> defaultClosure) {
        super();
        iPredicates = clone ? FunctorUtils.copy(predicates) : predicates;
        iClosures = clone ? FunctorUtils.copy(closures) : closures;
`-`µaaaaaµ`iDefault = (Closure<? super E>) (defaultClosure `
`-`µaaaaaaaaaµ`== null ? NOPClosure.<E>nopClosure() : `
`-`µaaaaaaaaaµ`defaultClosure);`
£+£¤aaaaa¤£this.iDefault = (defaultClosure == null ?£
£+£¤aaaaaaaaa¤£NOPClosure.nopClosure() : defaultClosure);£
    }
\end{lstlisting}
\autoref{lst:switch-closure} shows an excerpt of the original sources for \texttt{org.\allowbreak{}apache.\allowbreak{}commons.\allowbreak{}collections4.\allowbreak{}functors.\allowbreak{}SwitchClosure}.  The line highlighted in red is the original line. The line highlighted in green is the corresponding line as decompiled by \procyon, \cfr and \jd. None of them is able to correctly reproduce the cast to \texttt{Closure<? super E>}. This leads to a compilation error as the method \texttt{NOPClosure.<E>nopClosure()} return type is \textit{Closure<E>}, which is not a subtype of \texttt{Closure<? super E>} in the general case.

As the decompiled sources for \texttt{SwitchClosure} produced by all decompilers contain at least one error on this constructor, no solution is completable with the \store at the end of \autoref{alg:meta-algo}. Therefore, the meta decompilation fails to produce recompilable sources.

\subsection{Discussion}

In this section we discuss how alternative design decisions might be applied for meta decompilation. 
In particular, we discuss the use of different oracles to choose among decompiled fragments and the order of decompilers in \autoref{alg:meta-algo}.

A benefit of embedding a compiler in the meta decompiler is that it allows to use many different oracles to pick among the decompiled (and optionally recompiled) fragments. In this paper, we use the syntactic correctness assessment done by the compiler. But it would be possible to use other oracles. For certain decompilation use cases, such as source recovery, tests covering the original bytecode could be available. In the case of reverse engineering, it is realistic to assume  that one has access to a set of inputs with known outputs for the reversed program. They may be used as a test suite. The challenge with this approach is the granularity of the oracle provided by test cases. For meta decompilation to work, the granularity of the oracle needs to be finer or equal to the one of transplantation. In this work, we use \ecj errors as we are able to map them to code fragments. This allows us to label a fragment as correct and incorrect. 

Another oracle can be based on the bytecode distance between the decompiled-then-recompiled fragment and its original bytecode counterpart. This could be considered as a heuristic to minimize the likelihood of semantic differences between both fragments. In this work we measure bytecode distance with \texttt{JarDiff}, but \texttt{SootDiff}~\cite{sootdiff} could also be used, as its authors announce that it tolerates some control flow graph equivalent transformation.

Furthermore, depending on the metric that a decompiler user favours, the order of the decompilers used through meta decompilation may  change. In this work we rank decompilers according to the number of classes for which they produce semantically equivalent modulo inputs sources. If a user favours the rate of deceptive decompilation to be as low as possible, \cfr could be put first. Inverting the order of \procyon and \cfr for \metadc, on the $157$ classes presented in \autoref{sec:rq5}, yields only $38$ decompiled classes that are semantically equivalent modulo inputs. But it produces only $4$ deceptive decompilations.

\begin{mdframed}[style=mpdframe]
\textbf{Highlights about meta-decompilation:} 
To summarize, we have devised and implemented a novel approach to merge results from different decompilers, called meta-decompilation. This tool  handles $59$ of the $157$ cases ($37.6\%$) previously not handled by any decompiler. Meta-decompilation is, to our knowledge, a radically new idea that has never been explored before. Our experiments demonstrate the feasibility and effectiveness of the idea.
\end{mdframed}
\section{Threats to Validity}\label{sec:threats}
In this section, we report about internal, external and reliability  threats against the validity of our results. 

\paragraph{Internal validity.}

The internal threats are related to the metrics employed, especially those used to compare the syntactic distortion and semantic equivalence modulo inputs between the original and decompiled source code. Moreover, the coverage and quality of the test suite of the projects under study influences our observations about the semantic equivalence of the decompiled bytecode. To mitigate this threat, we select a set of mature open-source projects with good test suites as study subjects, and rely on state-of-the-art AST and bytecode differencing tools.

\paragraph{External validity.}
The external threats refer to what extent the results obtained with the studied decompilers can be generalized to other Java projects. To mitigate this threat, we reuse an existing dataset of Java programs which we believe is representative of the Java world. Moreover, we added a handmade project which is a collection of classes used in previous decompilers evaluations as a baseline for further comparisons.\looseness=-1

\paragraph{Reliability validity.} Our results are reproducible, the experimental pipeline presented in this study is publicly available online. We provide all necessary code to replicate our analysis, including AST metric calculations and statistical analysis via R notebooks.\footnote{\url{https://github.com/castor-software/decompilercmp/tree/master/notebooks}}\looseness=-1
\section{Related work}\label{sec:related}

This paper is related to previous works on bytecode analysis, decompilation and program transformations. In this section, we present the related work on Java bytecode decompilers along these lines.\looseness=-1

Kerbedroid \cite{jang2019kerberoid} is the closest related work. The work focuses on decompilers for Android and starts from the same observation as ours: decompilers perform differently with varying applications due to the various strategies to handle information lost in compilation. Kerbedroid is a meta-decompiler that stitches together results from multiple decompilers. Our current work shares the same observation, while contributing two key novel points. We perform an in-depth assessment of the different strategies implemented in 8 decompilers, with respect to three quality attributes, including equivalence modulo-input to compare the behavior of decompiled bytecode.
\metadc leverages the partial results from 8  decompilers instead of 3 for Kerbedroid, which increases the coverage of various corner cases in the bytecode.

The evaluation of decompilers is closely related to the assessment of compilers. In particular, Le et al.~\cite{Le2014} introduce the concept of semantic equivalence modulo inputs to validate compilers by analyzing the interplay between dynamic execution on a subset of inputs and statically compiling a program to work on all kind of inputs. 
Blackburn et al.~\cite{dacapo} propose a set of benchmarking selection and evaluation methodologies, and introduces the DaCapo benchmarks, a set of open source, client-side Java benchmarks.
Naeem et al.~\cite{Naeem2007} propose a set of software quality metrics aimed at measuring the effectiveness of decompilers and obfuscators. In 2009, Hamilton et al.~\cite{Hamilton2009} show that decompilation is possible for Java, though not perfect. In 2017, Kostelansky et al.~\cite{Kostelansky2017} perform a similar study on updated decompilers. In 2018, Gusarovs~\cite{Gusarovs2018a} performed a study on five Java decompilers by analyzing their performance according to different handcrafted test cases. All those works demonstrate that Java bytecode decompilation is far from perfect.\looseness=-1

Decompilers are disassemblers are closely related, and each pair of binary format, target language poses specific challenges (see Vinciguerra et al.   \cite{Vinciguerra2003} for C++, Khadra et al. \cite{Khadra2016} for ThumbISA, Grech et al. \cite{grech2019gigahorse} for Ethereum bytecode). 
With dissassembling, types must be reconstructing \cite{Troshina2010}, as well as assignment chains \cite{Emmerik2007}.  As we do in this paper, some researchers focus on reassembling disassembled binary code \cite{wang2015, navid2019, montoya2019}.

A recent trend in decompilation is to use neural networks \cite{Katz2018,li2019adabot,fu2019coda}. For example, Katz et al.~\cite{Katz2018} present a technique for decompiling binary code snippets using a model based on Recurrent Neural Networks, which produces source code that is more similar to human-written code and therefore more easy for humans to understand. This a remarkable attempt at driving decompilation towards a specific goal. Lacomis et al.~\cite{lacomis2019dire} propose a probabilistic technique for variable name recovery. Schulte et al.~\cite{Schulte2018} use evolutionary search to improve and recombine a large population of candidate decompilations by applying source-to-source transformations gathered from a database of human-written sources. Miller and colleagues \cite{miller2019probabilistic} model the uncertainty  due to the information loss during compilation  using probabilities and propose a novel disassembly technique, which computes a probability for each address in the code space, indicating its likelihood  of being a true positive instruction.

\section{Conclusion}\label{sec:conclusion}

In this work, we presented a fully automated pipeline to assess  Java bytecode decompilers with respect to their capacity to produce compilable,  equivalent modulo-input, and readable code. 
We assessed eight decompilers with a set of $2041$ classes from $14$ open-source projects compiled with two different compilers. 
The results of our analysis show that bytecode decompilation is a non-trivial task that still requires human work. Indeed, even the highest ranking decompiler in this study  produces syntactically correct output for $84\%$ of classes of our dataset and semantically equivalent modulo inputs output for $78\%$.
We extract $157$ classes for which no decompiler produces semantically equivalent sources. These classes illustrate how generics and, in particular, generic with wildcards and type bounds are challenging for all decompilers. Yet the Java language with its diversity of compilers and versions makes room for many corner cases that require extensive testing and development effort from decompilers authors.
Meanwhile, the diversity of implementation of these decompilers allows to merge their different results to bypass the shortcomings of single decompilers. We called this approach `meta decompilation` and implemented it in a tool called \metadc. Our experimental results show that \metadc can produce \semi sources for $37.6\%$ of classes for which, previously, no single decompiler could.

\section*{Acknowledgments}\label{sec:ak} This work has been partially supported by the Wallenberg Autonomous Systems and Software Program (WASP) funded by Knut and Alice Wallenberg Foundation and by the TrustFull project funded by the Swedish Foundation for Strategic Research.\looseness=-1

\end{document}